\documentclass{aa}  

\usepackage{graphicx}
\usepackage{txfonts}
\usepackage{booktabs}
\usepackage{multirow}
\usepackage[colorlinks=true, linkcolor=blue, citecolor=blue, filecolor=blue,urlcolor=blue]{hyperref}

\begin{document}

   \title{Expected insights into Type Ia supernovae from LISA's gravitational wave observations}

   \author{Valeriya Korol
          \inst{1}%\fnmsep\thanks{Just to show the usage
          %of the elements in the author field}
          \and
          Riccardo Buscicchio \inst{2,3}
          \and 
          Ruediger Pakmor \inst{1}
          \and
          Javier Mor\'an-Fraile \inst{4}
          \and
          Christopher J. Moore\inst{5,6,7}
          \and
          Selma E. de Mink\inst{1}
          }
          %Stephen Justham \inst{1}
          %\and
          %Selma de Mink \inst{1}%\fnmsep\thanks{Just to show the usage
          %of the elements in the author field}

   \institute{Max-Planck-Institut f{\"u}r Astrophysik, Karl-Schwarzschild-Stra{\ss}e 1, 85748 Garching, Germany\\
              \email{korol@mpa-garching.mpg.de}
         \and
             Dipartimento di Fisica “G. Occhialini”, Università degli Studi
             di Milano-Bicocca, Piazza della Scienza 3, 20126 Milano, Italy
        \and 
            INFN, Sezione di Milano-Bicocca, Piazza della Scienza 3, 20126 Milano, Italy
        \and
            Heidelberger Institut für Theoretische Studien (HITS),
            Schloss-Wolfsbrunnenweg 35, 69118 Heidelberg, Germany 
             %\email{other@blabla.com}
        \and
        Institute of Astronomy, University of Cambridge, Madingley Road, Cambridge, CB3 0HA, UK
        \and
        Kavli Institute for Cosmology, University of Cambridge, Madingley Road, Cambridge, CB3 0HA, UK
        \and
        Department of Applied Mathematics and Theoretical Physics, Centre for Mathematical Sciences, University of Cambridge, Wilberforce Road, CB3 0WA, UK
             }

   \date{Received 4 July 4 2024/ Accepted 10 September 2024}

% \abstract{}{}{}{}{} 
% 5 {} token are mandatory
 
  \abstract
   {The nature of progenitors of Type Ia supernovae has long been debated, primarily due to the elusiveness of the progenitor systems to traditional electromagnetic observation methods. We argue that gravitational wave observations with the upcoming Laser Interferometer Space Antenna (LISA) offer the most promising way to test one of the leading progenitor scenarios — the double-degenerate scenario, which involves a binary system of two white dwarf stars.
   In this study we review published results, supplementing them with additional calculations for the context of Type Ia supernovae.
   We discuss the fact that LISA will be able to provide a complete sample of double white dwarf Type Ia supernova progenitors with orbital periods shorter than 16--11 minutes (gravitational wave frequencies above 2--3 millihertz). Such a sample will enable a statistical validation of the double-degenerate scenario by simply counting whether LISA detects enough double white dwarf binaries to account for the measured Type Ia merger rate in Milky Way-like galaxies. Additionally, we illustrate how LISA’s capability to measure the chirp mass will set lower bounds on the primary mass, revealing whether detected double white dwarf binaries will eventually end up as a Type Ia supernova. We estimate that the expected LISA constraints on the Type Ia merger rate for the Milky Way will be 4-9\%. We also discuss the potential gravitational wave signal from a Type Ia supernova assuming a double-detonation mechanism and explore how multi-messenger observations could significantly advance our understanding of these transient phenomena. }

  % {We aim to outline what will we learn about Type Ia supernovae from LISA's gravitational wave perspective.}
   %{Miscellaneous.}
   %{We show ... }


   \maketitle
%
%-------------------------------------------------------------------

\section{Introduction}

Type Ia supernovae (SNe~Ia) are amongst the most energetic explosions in the Universe, producing a luminosity of about $10^{43}$\,erg\,s$^{-1}$ near maximum light. They emerged as a distinct class based on spectral signatures: the absence of hydrogen and helium and the presence of broad features of silicon, calcium, and iron (e.g. see the review by \citealt{fil97}). Notably, the light curve shapes of many SNe~Ia closely resemble each other: peaking within 15–20 days and then declining rapidly. As a consequence, these standardisable (also referred to as `normal') SNe~Ia have been adopted as cosmological standard candles and provided the first evidence for the acceleration of cosmic expansion \citep{rie98,per99}. The luminosity that characterises the light curve of a (normal) SN~Ia originates from the radioactive decay of $^{56}$Ni to $^{56}$Co and then to stable $^{56}$Fe. The energetics and chemical composition imply that these SNe result from the thermonuclear combustion of a white dwarf \citep{hoy60}. Despite SNe~Ia being among the most studied transient phenomena, there is still no consensus on other fundamental aspects, such as the nature of the progenitor and explosion mechanisms, both from theoretical and observational perspectives (see recent reviews by \citealt{mao14}, \citealt{rui20}, and \citealt{liu23}).

There is a broad consensus that the white dwarf explosion in a SN~Ia is triggered by the interaction with a companion star. Identifying the nature of the companion (and the nature of the explosion), however, has been a long-standing challenge \citep[e.g.][]{liv18}. 
Historically, the field has developed around two main progenitor scenarios: single-degenerate and double-degenerate\footnote{Here `degenerate' refers to white dwarfs, which are supported by the pressure of degenerate electrons, in contrast to normal (non-degenerate) stars, which are supported by thermal pressure.}. In the single-degenerate scenario, a white dwarf accretes matter from a non-degenerate companion (e.g. a main sequence star, sub-giant, or helium star) until it approaches the Chandrasekhar mass limit and explodes \citep{whe73, non84}. In contrast, in the double-degenerate scenario, the companion is another white dwarf, and the two white dwarfs are brought together to interact or merge directly via gravitational wave (GW) radiation \citep{whe73, ibe84}.  We refer the reader to the recent review by \citet{sok24} for a more articulated picture of SN~Ia progenitor scenarios. Each scenario offers a plausible explanation, and in the absence of definitive evidence, it is difficult to conclusively favour one over the other \citep[e.g.][]{mao14}. 

%%%%%%%%%%%%%%%%%%%%%%%%%%%%%%%%%%%%%%
  \begin{figure*}
   \centering
     \includegraphics[width=9cm]{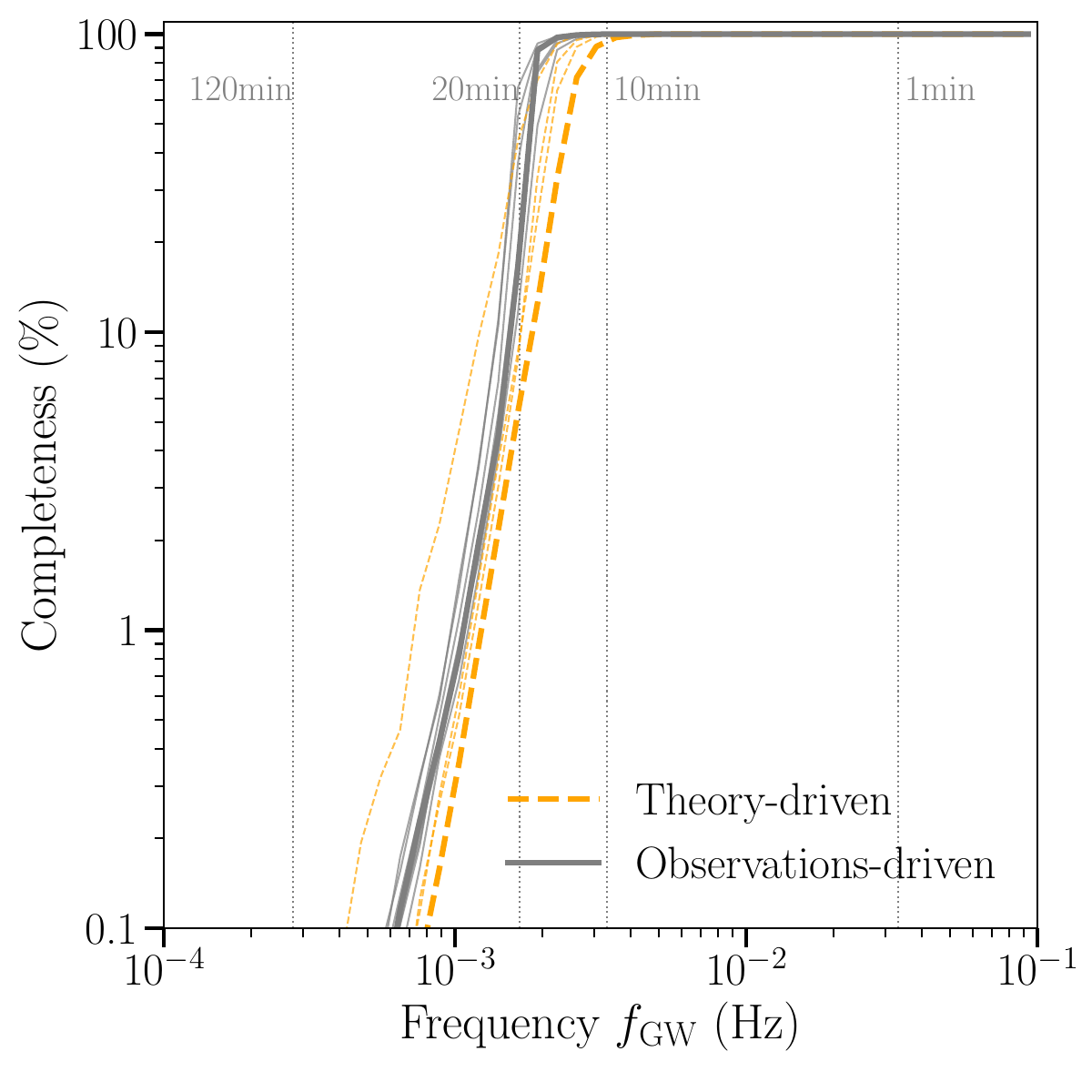}
      \includegraphics[width=9cm]{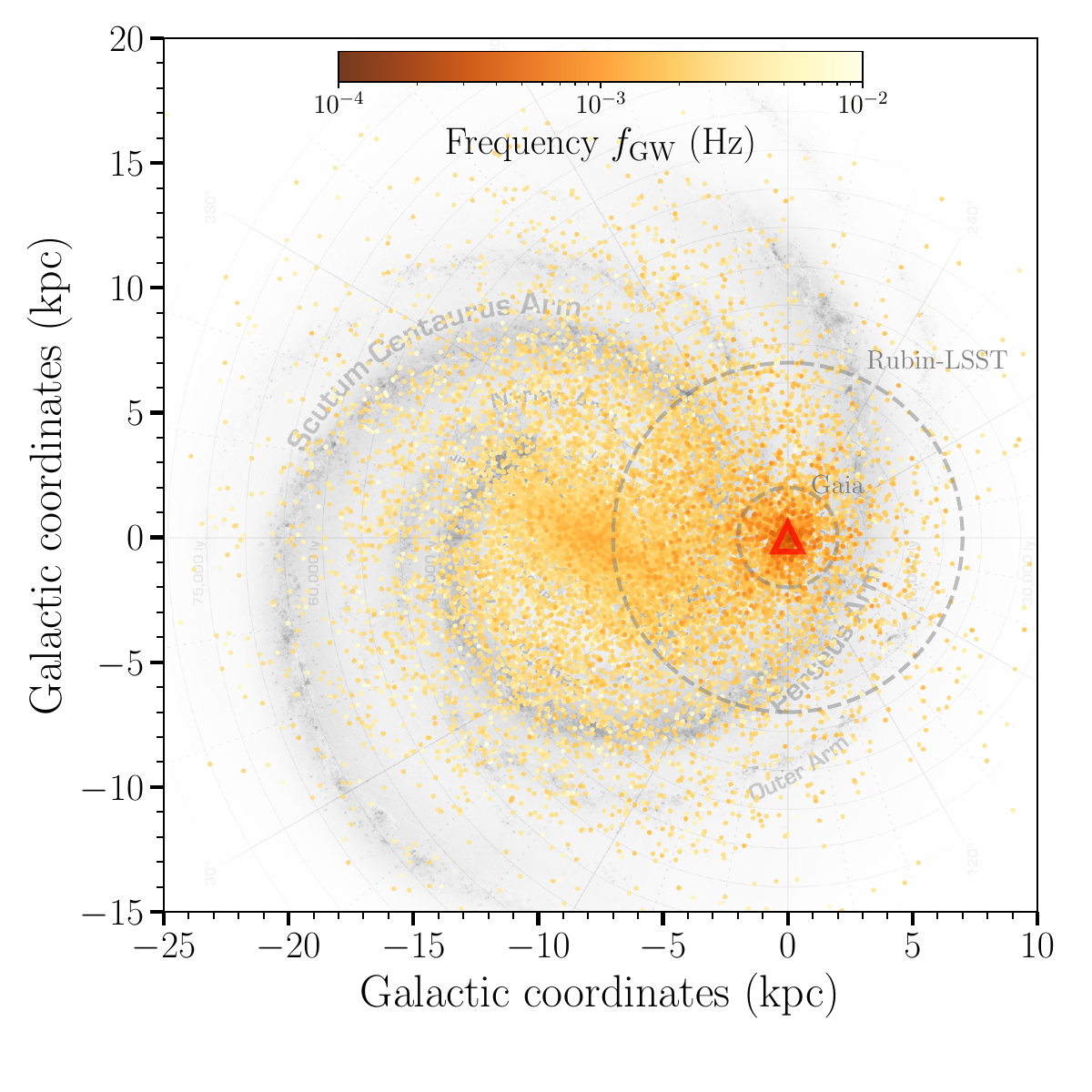}
      \caption{LISA's capability to probe the Galactic WD+WD binary population. {\it Left panel:} Completeness of LISA's Galactic WD+WD sample as a function of frequency estimated based on sets of mock Galactic catalogues: theory-driven (dashed orange lines; \citealt{kor17, wil21}) and observation-driven (solid grey lines; \citealt{kor22}). Thick lines represent the respective default models, and thin lines model variations within each set (e.g. different prescriptions for binary interactions in the theory-driven models).
      Vertical dotted lines mark the respective binary orbital periods. {\it Right panel:} Spatial distribution of binaries with $\rho_{\rm 1yr}\geq7$ within the Galaxy adapted from \citet{wil21}, colour-coded by GW frequency. In the background we show an artist's impression of our current view of the Milky Way. The red triangle at $(0,0)$ shows LISA's position, while the Galactic Centre is at $(-8.2,0)$. For comparison, we show estimated observational horizons for WD+WD binaries for {\it Gaia} and {\it Rubin}-LSST surveys \citep{kor17}. } 
    \label{fig:LISA_completeness}
   \end{figure*}
%%%%%%%%%%%%%%%%%%%%%%%%%%%%%%%%%%%

There are significant challenges associated with detecting double-degenerate SN Ia progenitors through electromagnetic (EM) observations, primarily due to the intrinsic faintness of these binaries. So far very few binaries have been identified as potential double-degenerate SN~Ia progenitors \citep[see e.g.][for a recent overview]{mun24}.
\citet{reb19} highlighted that even with next-generation large-aperture telescopes, such as the Extremely Large Telescope (ELT), the estimated probability of identifying and validating SN~Ia progenitors  (by measuring the binary components' masses) will remain remarkably low. 
Here we discuss why GW observations with the Laser Interferometer Space Antenna (LISA) offer a promising alternative for determining the nature of SN~Ia progenitor systems, overcoming the limitations faced by traditional EM observation methods.

In this study we explore the SN~Ia progenitor question from the perspective of future GW observations. 
Specifically, we discuss the potential of LISA, an upcoming millihertz GW observatory that has recently been scheduled for launch by the European Space Agency in the mid-2030s \citep{Redbook}. Our arguments also hold for other space-based GW observatories such, as TianQin \citep{TianQin, hua20}, Taiji \citep{Taiji}, and the Lunar Gravitational Wave Antenna \citep{LGWA,bra23,LGWAastro}.
LISA's sensitivity to the shortest-period (less than about 2 hours) double white dwarf (WD+WD) binaries  in the Milky Way and its nearest satellite galaxies presents a unique opportunity to test the double-degenerate scenario directly \citep{astrowp}. 

The structure of this paper is as follows. In Sect.~\ref{sec:2} we review the constraints on double-degenerate WD+WD progenitors in the Milky Way that we will likely be able to place based on LISA observations  \citep{Redbook}, including the incidence of these binaries (Sect.~\ref{sec:complete_sample}), chirp mass constraints (Sect.~\ref{sec:Mchirp}), and their overall merger rate (Sect.~\ref{sec:merger_rate}). In Sect.~\ref{sec:3} we discuss the fortunate possibility of directly observing the final inspiral and a SN~Ia explosion during LISA's operation time. We adopt a 3D hydrodynamical simulation, in which the primary white dwarf undergoes the double detonation mechanism, to illustrate the evolution of GW strain in frequency over the last few orbits prior to a SN~Ia event (Sect.~\ref{sec:GW_signal}). We also discuss the implications of multi-messenger observations (Sect.~\ref{sec:multi-messenger}), including expectations for neutrino signals (Sect.~\ref{sec:neutrinos}). Finally, we present a summary and the conclusions of our study in Sect.~\ref{sec:4}.

%--------------------------------------------------------------------
\section{Constraints on the double-degenerate progenitor systems in the Milky Way} \label{sec:2}

%%%%%%%%%%%%%%%%%%%%%%%%%%%%%%%%%%%%%%
        \begin{figure*}
   \centering
   \includegraphics[height=8cm]{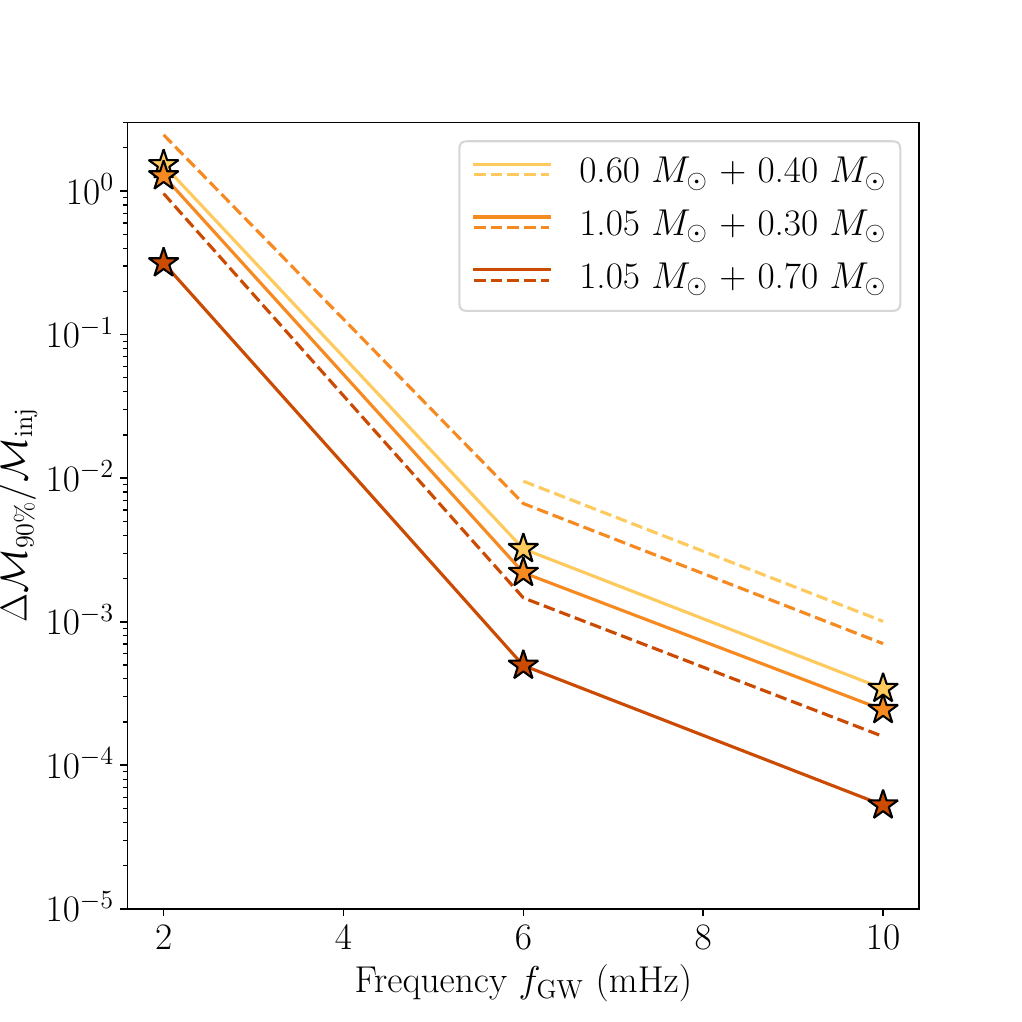}
\includegraphics[height=7.25cm]{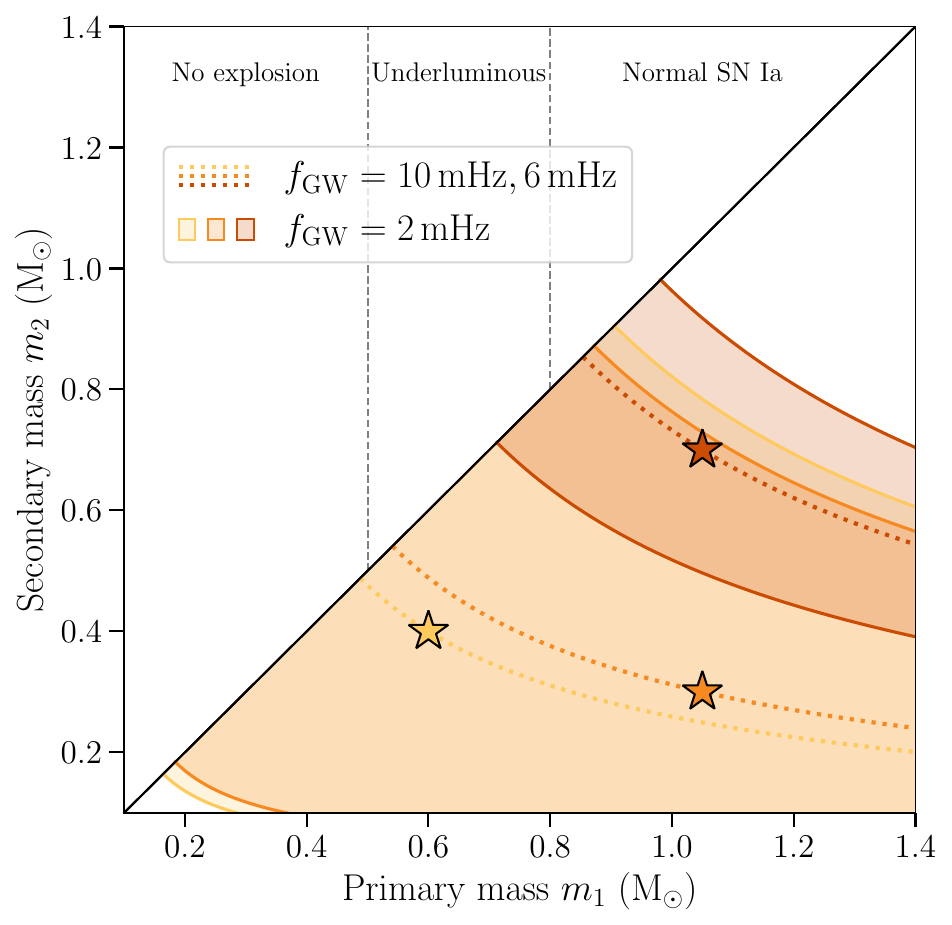}
      \caption{LISA's capability to measure chirp masses  and constrain binary component masses. {\it Left panel}: Fractional uncertainty on the chirp mass.\ Solid (dashed) lines represent face-on (edge-on) binary configurations, and stars denote representative simulated systems. {\it Right panel}: Constraints on component masses.\ The shaded areas between solid lines denote sources at 2\,mHz. Dotted lines represent sources at 6 and 10\,mHz. These constraints assume that the symmetric mass ratio (i.e. $m_1m_2/(m_1+m_2)^2$) is unknown, and they are sampled from a uniform prior.  Vertical dashed grey lines delimit three representative classes: configurations that produce no explosion ($m_1<0.5\,$M$_\odot$), those likely resulting in a low-luminosity thermonuclear transient ($0.5 < m_1 <0.8\,$M$_\odot$), and those that will likely result in a normal SN~Ia ($m_1>0.8\,$M$_\odot$). 
      }
      \label{fig:Mchirp}
   \end{figure*}
%%%%%%%%%%%%%%%%%%%%%%%%%%%%%%%%%%%%%

The dense and compact nature of white dwarf stars, while challenging for EM observations, offers advantages for the GW detection. In fact, in the following we argue that the best way of surveying double-degenerate SN~Ia progenitors is via GW observations with a mission like LISA.

GW radiation from a circular WD+WD binaries can be modelled  with a nearly monochromatic sinusoidal waveform \citep[e.g.][]{WaveformsWP}. Such a waveform can be described by eight key parameters: the GW amplitude $\mathcal{A}$, GW frequency $f_{\mathrm{GW}}$, its derivative $\dot{f}_{\mathrm{GW}}$, the binary's sky coordinates ($\lambda, \beta$), inclination angle $\iota$, polarisation angle $\psi$, and the initial phase $\phi_0$. For circular binaries, the GW frequency is exactly $f_{\mathrm{GW}} = 2/P_{\mathrm{orb}}$, with $P_{\mathrm{orb}}$ being the binary's orbital period. The strain amplitude of the signal is given by
\begin{equation} \label{eq:amp}
\mathcal{A} = \frac{2 (G \mathcal{M})^{5/3}}{c^4 d} (\pi f_{\mathrm{GW}})^{2/3},
\end{equation}
and is determined by the binary's GW frequency, distance $d$ and chirp mass $\mathcal{M}={(m_1 m_2)^{3/5}}/{(m_1 + m_2)^{1/5}}$, for component masses $m_1\geq m_2$; $G$ and $c$ are respectively the gravitational constant and speed of light. Equation~\ref{eq:amp} reveals that binaries characterised by higher frequencies and larger masses produce GWs with larger amplitude, yielding signals with higher signal-to-noise ratios:
\begin{equation}
    \rho \propto {\cal A} \sqrt{\frac{T_{\rm obs}}{S_n(f_{\rm GW})}},
\end{equation}
where $T_{\rm obs}$ is the observation time and $S_{\rm n}(f_{\rm GW})$ is the noise power spectral density, a quantity describing the noise of the LISA detector as a function of frequency \citep[e.g.][]{moo15}, at the binary's frequency \citep[for the full derivation, see Appendix A of][]{fin23}. 
We note that, unlike EM observations — which are based on the measurement of energy flux and scale with $1/d^2$ — GW observations offer direct measurements of the signal's amplitude and thus scale with $1/d$. Consequently, GW observations enable the detection of EM-dim sources (or even entirely invisible dormant binary black holes) at distances that are not directly accessible with EM observations \citep[e.g.][]{kor17,ses20}.

\subsection{Probing the number of progenitor systems} \label{sec:complete_sample}

To study the expected completeness of LISA's WD+WD sample, we utilised published mock catalogues assembled based on both theory-driven \citep{kor17,wil21} and observations-driven \citep{kor22} population synthesis.
Our theory-driven models are based on the \textsc{SeBa} stellar and binary evolution module \citep{spz96,nel01,too12} with variations in the underlying assumptions for the binary interactions \citep[for details see][]{too12}.
While our observations-driven models are based on a statistical method developed  to characterise the binarity of large spectroscopic samples of white dwarfs with sparsely sampled radial velocity data \citep[][]{badenes12,maoz12,mao18}. Variations in our observations-driven models are due to different underlying assumptions for the binary fraction, WD+WD separation (at formation), and mass distributions  \citep[for details see][]{kor22}.

We assessed the completeness as the percentage of detectable binaries relative to the total underlying WD+WD population as a function of frequency. 
Importantly, the populations selected for this study are directly comparable, as they were analysed under the same mission specifications, including the noise model, a 4-year mission duration, and a detection threshold of $\rho_{4{\rm yr}}=7$. This analysis employed the same pipeline as \citet{kar21}, which self-consistently estimates the unresolved stochastic foreground from the input population and resolves binaries with a signal-to-noise ratio above the threshold. The result is presented in the left panel of Fig.~\ref{fig:LISA_completeness}, where solid grey lines represent the family of observations-driven models and orange dashed lines represent the family of theory-driven models. 
In the right panel, we show the example of the spatial distribution within the Milky Way of WD+WD binaries with $\rho_{\rm 1yr}\geq7$ from \citealt{wil21}, colour-coded by GW frequency. It is evident that binaries emitting at $f_{\rm GW}\geq 2-3\,$mHz (in pale orange to yellow) can be detected throughout the entire Galaxy. Although not strictly comparable due to differences in the WD+WD detection analysis, we expect that other recently published models will show similar results \citep[e.g.][]{lam19,bre20,li20,tie23,li23,tang24}

Figure~\ref{fig:LISA_completeness} illustrates that LISA is expected to achieve a complete sample of WD+WD binaries with GW frequencies above $2-3\,$mHz, corresponding to orbital periods shorter than 16 to 11 minutes. Thus, based on the LISA sample, we would be able to quantitatively assess the contribution of double-degenerate progenitors to the measured SN~Ia merger rate from EM observations \citep[see also][]{Redbook}. This could be achieved by simply counting whether LISA detects enough WD+WD binaries, importantly without concerns regarding selection effects.

%--------------------------------------------------------------------
\subsection{Probing the masses of progenitor system} \label{sec:Mchirp}

To identify potential SN~Ia progenitors within LISA's sample, constraints on the binary components' masses are essential, particularly considering that only a subset of the most massive WD+WD binaries (with primary mass $m_1 > 0.8$\,M$_\odot$) are more likely to lead to SN~Ia events. This is an essential requirement because lower-mass white dwarfs may not produce significant amounts of $^{56}$Ni, even if they undergo a thermonuclear explosion  \citep[e.g.][]{sim10}. We also assumed that binaries $0.5 < m_1 < 0.8$\,M$_\odot$ result in thermonuclear explosions of lower luminosity than a normal SN~Ia, while those with $m_1 < 0.5$\,M$_\odot$ fail to detonate \citep[e.g.][]{mor24,shen24}. 

At GW frequencies greater than 2-3\,mHz, LISA is expected to measure not only the GW frequency and amplitude of WD+WD signals but also the frequency derivative -- the so-called chirp -- resulting from GW radiation \citep[e.g.][]{Redbook}. Through the measurement of $f_{\mathrm{GW}}$ and its derivative $\dot{f}_{\mathrm{GW}}$, it becomes possible to deduce the binary system's chirp mass. Under the assumptions that the binary is detached, non-interacting, and on a circular orbit, which are expected to hold for most of the LISA-detectable WD+WD population, the GW frequency derivative can be expressed as\begin{equation}
\label{eqn:fdot}
   \dot{f}_{\rm GW} = \frac{96}{5}\frac{(G\mathcal{M})^{5/3}}{\pi c^5} (\pi f_{\rm GW})^{11/3}.
\end{equation}
We note that the above expression ignores tidal effects, which could become important at higher frequencies for example, binaries considered below \citep[e.g.][]{sha14,wol21,tou24}. We also note that ignoring these effects, especially at $f_{\rm GW} > 10\,$mHz, can lead to non-negligible biases in the mass estimate \citep{fia24}.

To showcase LISA's potential for measuring the chirp mass of double-degenerate SN~Ia progenitor systems, we considered three examples: 0.6\,M$_\odot$+0.4\,M$_\odot$, 1.05\,M$_\odot$+0.3\,M$_\odot$, and 1.05\,M$_\odot$+0.7\,M$_\odot$. These combinations yield chirp masses of 0.42\,M$_\odot$, 0.47\,M$_\odot$, and 0.74\,M$_\odot$, respectively. The binary with the lowest chirp mass is representative of a potential progenitor for thermonuclear transients of lower luminosity than a normal SN~Ia, whereas the latter two configurations are representative of a normal SN~Ia progenitor \citep[e.g.][]{shen15}.
We generated GW signals and performed parameter estimations for these three example binaries within a Bayesian inference framework using \textsc{}{\textsc{Balrog}} \citep{2019PhRvD.100h4041B,roe20,bus21,kle22,fin23}.
We simulated signals with initial frequencies of 2, 6, and 10\,mHz, while bracketing dependence on the binaries' orientation with face-on and edge-on configurations. All sources are (arbitrarily) located in the Galactic centre, assuming a distance of 8.2\,kpc \citep[e.g.][]{gravity21}. 

In Fig.~\ref{fig:Mchirp} we showcase LISA's expected capability to measure the binary's chirp mass across different frequencies (left panel) and the corresponding constraints on the component masses (right panel).  
Our results reveal that at 2\,mHz, the chirp mass measurement is constraining only for the most massive case among the considered examples.
As the frequency increases beyond 4-5 mHz, the precision of the chirp mass measurements significantly improves for all three examples, dropping rapidly to 0.01\% - 0.001\% at 10\,mHz. From the right panel, we observe that at 2 mHz (as indicated by coloured bands), the chirp mass measurement implies a minimum primary mass of 0.2\,M$_\odot$ for the binaries with chirp masses of 0.42\,M$_\odot$ and the 0.47\,M$_\odot$, and a minimum mass of 0.7\,M$_\odot$ for the one with a 0.74\,M$_\odot$ chirp mass. Consequently, for the latter example, we can predict an eventual SN~Ia with a posterior probability of 96\%, estimated by assuming a uniform prior in mass ratio and by counting posterior samples within the $m_1>0.8\,$M$_\odot$ region.
As the frequency increases to 6\,mHz and higher, the uncertainty in chirp mass measurement reduces to less than 1\%; as a consequence, constraints on components' mass converge to a line. This precision translates to a lower bound for the primary mass of 0.49 M$_\odot$ for the 0.42 M$_\odot$ chirp mass binary, and 0.55 M$_\odot$ for the 0.47 M$_\odot$ chirp mass binary. For these two cases, the probability of eventually resulting in a thermonuclear transient is 100\%, estimated by assuming a uniform prior in mass ratio and by counting posterior samples within the $m_1>0.5\,$M$_\odot$ region.
For the most massive of the considered examples, the lower limit on the primary mass can be set to 0.87\,M$_\odot$. This implies that LISA will be able to assign a 100\% probability that this binary is a progenitor of a normal SN~Ia (i.e. $m_1>0.8\,$M$_\odot$).

\begin{table*} 
\caption{Point estimates from simulations with \textsc{Balrog}.}
\resizebox{\textwidth}{!}{%
\begin{tabular}{ccccccc|cccc|ccc}
\toprule
\toprule
\multicolumn{7}{c|}{ Injection parameters} & \multicolumn{4}{c|}{Posterior} & \multicolumn{3}{c}{Classification}\\
\midrule
$\mathcal{M}^{\rm inj}$ & $m_1^{\rm inj}$ & $m_2^{\rm inj}$ & $f^{\rm inj}$ & $\tau^{\rm inj}$ & $\cos\iota^{\rm inj}$ & S/N & $\Delta \mathcal{M}^{90\%}$ & $\Delta \tau^{90\%} $ & $\Delta \mathcal{M}^{90\%} / \mathcal{M}^{\rm inj}$ & $\Delta \tau^{90\%} / \tau^{\rm inj}$ & No expl. & Underl. & Normal\\
$[\mathrm{M}_\odot]$ & $[\mathrm{M}_\odot]$ & $[\mathrm{M}_\odot]$ & $[{\rm Hz}]$ & $[{\rm Myr}]$ & & & $[\mathrm{M}_\odot]$ & $[{\rm Myr}]$ &  & & $[\%]$ & $[\%]$& $[\%]$\\
\midrule
\multirow{7.0}{*}{0.42} & \multirow{7.0}{*}{0.60} & \multirow{7.0}{*}{0.40} & \multirow{2}{*}{0.002} & \multirow{2}{*}{1.345} & 0 & 5.1 & 5.57312 & 0.351 & 13.1211 & 0.2607 & 16.0 & 25.9 & 58.0 \\
&  &  &  &  & 1 & 15.0 & 0.64505 & 7.764 & 1.5187 & 5.7705 & 21.7 & 31.7 & 46.6 \\
\cmidrule[0.5pt]{4-14}
  &  &  & \multirow{2}{*}{0.006} & \multirow{2}{*}{ 0.072 } & 0 & 57.4 & 0.00404 & 0.0011 & 0.0095 & 0.0159 & 3.4 & 56.0 & 40.6 \\
  &  &  &  &  & 1 & 169.8 & 0.00137 & 0.0004 & 0.0032 & 0.0054 &  3.0 & 54.8 & 42.2\\
\cmidrule[0.5pt]{4-14}
  &  &  & \multirow{2}{*}{0.010} & \multirow{2}{*}{0.018} & 0 & 82.0 & 0.00043 & 0.000031 & 0.0010 & 0.0017 & 3.1 & 55.3 & 41.6 \\
  &  &  &  &  & 1 & 243.1 & 0.00015 & 0.00001 & 0.0003 & 0.0006 &  3.8 & 56.6 & 39.6\\
\cmidrule[0.5pt]{1-14}
 \multirow{7.0}{*}{0.47} & \multirow{7.0}{*}{1.05} & \multirow{7.0}{*}{0.30} & \multirow{2}{*}{0.002} & \multirow{2}{*}{1.133} & 0 & 6.0 & 1.16005 & 5.589 & 2.4635 & 4.9332 & 18.1 & 25.6 & 56.3\\
  &  &  &  &  & 1 & 17.8 & 0.59892 & 6.316 & 1.2719 & 5.5747 & 19.3 & 33.2 & 47.4 \\
\cmidrule[0.5pt]{4-14}
  &  &  & \multirow{2}{*}{0.006} & \multirow{2}{*}{ 0.061 } & 0 & 68.2 & 0.00313 & 0.0007 & 0.0067 & 0.0111 & 0.0        & 51.0  & 49.0\\
  &  &  &  &  & 1 & 201.7 & 0.00103 & 0.0002 & 0.0022 & 0.0037 &  0.0 & 51.2 & 48.8\\
\cmidrule[0.5pt]{4-14}
  &  &  & \multirow{2}{*}{0.010} & \multirow{2}{*}{0.015} & 0 & 97.4 & 0.00033 & 0.000018 & 0.0007 & 0.0012 & 0.0 & 49.8 & 50.2\\
  &  &  &  &  & 1 & 288.7 & 0.00011 & 0.000006 & 0.0002 & 0.0004 & 0.0 & 51.5 & 48.5 \\
\cmidrule[0.5pt]{1-14}
 \multirow{7.0}{*}{0.74} & \multirow{7.0}{*}{1.05} & \multirow{7.0}{*}{0.70} & \multirow{2}{*}{0.002} & \multirow{2}{*}{0.529} & 0 & 12.9 & 0.71376 & 1.613 & 0.9603 & 3.0476 & 6.2 & 19.0 & 74.8\\
  &  &  &  &  & 1 & 38.1 & 0.23422 & 0.296 & 0.3151 & 0.5597 & 0.0 & 4.5 & 95.5\\
\cmidrule[0.5pt]{4-14}
  &  &  & \multirow{2}{*}{0.006} & \multirow{2}{*}{ 0.028 } & 0 & 146.0 & 0.00109 & 0.00007 & 0.0015 & 0.0024 & 0.0 & 0.0 & 100.0 \\
  &  &  &  &  & 1 & 431.6 & 0.00037 & 0.00002 & 0.0005 & 0.0008 &  0.0 & 0.0 & 100.0\\
\cmidrule[0.5pt]{4-14}
  &  &  & \multirow{2}{*}{0.010} & \multirow{2}{*}{0.007} & 0 & 208.5 & 0.00012 & 0.000002 & 0.0002 & 0.0003 & 0.0 & 0.0 & 100.0\\
  &  &  &  &  & 1 & 617.8 & 0.00004 & 0.000001 & 0.0001 & 0.0001 & 0.0 & 0.0 & 100.0 \\
\bottomrule
\end{tabular}} \label{tab:1}
\tablefoot{The superscript `inj' stands for injected, i.e. assumed as the true value; the abbreviation `S/N' stands for the signal-to-noise ratio. The three leftmost columns represent classification probabilities for each source, according to the three classes defined in Sect.~\ref{sec:Mchirp}: `No explosion' ($m_1<0.5$\,M$_\odot$), `Underluminous' ($0.5\,{\rm M}_\odot < m_1 < 0.8\,{\rm M}_\odot$) and `Normal SN~Ia' ($m_1>0.8$\,M$_\odot$).}
\end{table*}

%--------------------------------------------------------------------
\subsection{Probing the merger rate of progenitor systems} \label{sec:merger_rate}

The measurement of the chirp mass offers a method for estimating the time until merger due to  GW radiation. The time to merger,  $\tau=t_{\rm merger}-t$, can be estimated, in a first post-Newtonian approximation, by integrating Eq.~\ref{eqn:fdot} \citep[e.g.][]{pet64}: 
\begin{equation} \label{eqn:tau}
    \tau(f_{\rm GW},{\cal M}) = \frac{5 c^5}{256 (G{\cal M})^{5/3}(\pi f_{\rm GW})^{8/3}},
\end{equation}
which is valid under the same assumptions as used in Eq.~\ref{eqn:fdot}.
Given that the GW frequency of WD+WD binaries is measured with high precision ($\Delta f_{\rm GW}/f_{\rm GW} \ll 10^{-5}$; e.g. \citealt{kar21,fin23,Redbook}), the uncertainty in $\tau$ primarily depends on LISA's ability to constrain the chirp mass, with $\Delta\tau/\tau \approx 5/3 \ \Delta{\cal M}/{\cal M}$. We provide the estimate of this measurement for the three considered examples in Table~\ref{tab:1}. 

Assuming that binaries in the LISA sample inspiral according to Eq.~\ref{eqn:tau}, their distribution in frequency will enable an estimate of the overall merger rate. 
To provide an order-of-magnitude estimate of LISA's ability to constrain the SN~Ia merger rate, ${\cal R}_{\rm SNIa}$, we considered a simplified scenario where all SN  ~Ia progenitors consist of binaries with masses of \(1.05\,\textrm{M}_{\odot} + 0.7\,\textrm{M}_{\odot}\) (corresponding chirp mass of \(0.74\,\textrm{M}_{\odot}\)). 
As illustrated in Sect.~\ref{sec:Mchirp}, binaries with similar chirp mass or higher can be identified as SN~Ia progenitors with high confidence, starting at 2\,mHz. 
Assuming that the WD+WD population is in a steady state, the spectral density of binaries resolved over the mission duration is given by 
\begin{equation} \label{eqn:dNdf}
\frac{\mathrm{d}N_{\rm SNIa}}{\mathrm{d}f_{\rm GW}} = \frac{5c^5 {\cal R}_{\rm SNIa} }{96 \pi^{8/3} (G{\cal M})^{5/3} f_{\rm GW}^{11/3}}.
\end{equation}
The rate can be recovered by integrating Eq. 5 over frequency and solving for ${\cal R}_{\rm SNIa}$:
\begin{equation}\label{eq:rate}
     {\cal R}_{\rm SNIa} = \frac{256 (G {\cal M})^{5/3}( \pi f_{\rm GW})^{8/3} }{5 c^5} N_{\rm SNIa}(>f_{\rm GW}) ,
\end{equation}
where $N_{\rm SNIa}(>f_{\rm GW})$ is the total number of binaries emitting above a certain frequency threshold $f_{\rm GW}$. From Eq.~\ref{eq:rate}, it follows that the error on  ${\cal R}_{\rm SNIa}$ depends on the errors in the number count of SN~Ia progenitor binaries and the error on the chirp mass. 
As discussed in Sect.~\ref{sec:complete_sample}, at frequencies higher than 2\,mHz, LISA can detect every relevant binary, allowing the estimate of the merger rate of \(1.05\,\textrm{M}_{\odot} + 0.7\,\textrm{M}_{\odot}\) without concerns about selection effects.

%%%%%%%%%%%%%%%%%%%%%%%%%%%%%%%%%

\begin{figure}
   \centering
   \includegraphics[height=8.5cm]{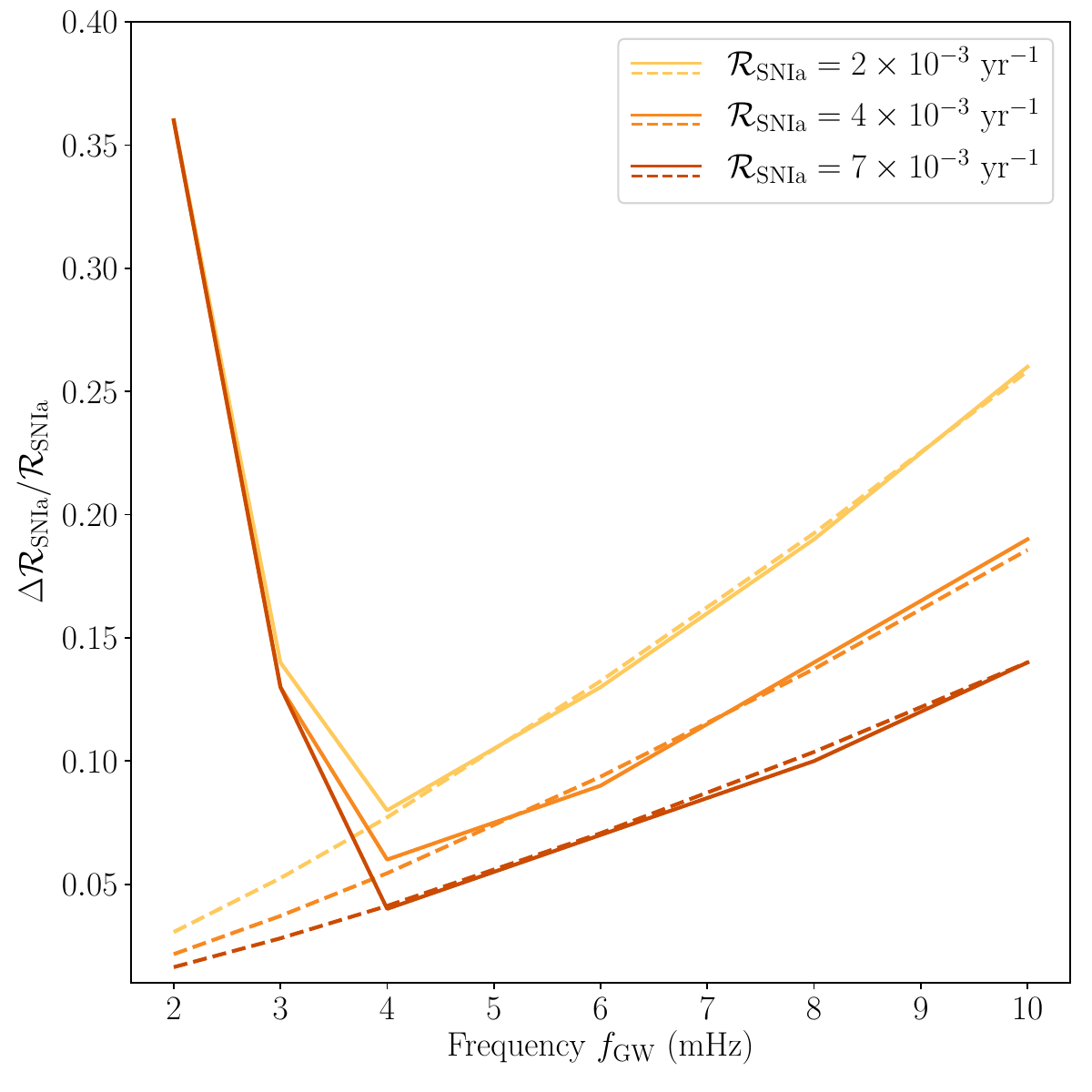}
      \caption{LISA’s capability to measure SN~Ia merger rates (${\cal R}_{\rm SNIa}$) from double-degenerate SN~Ia progenitors at frequencies between 2 and 10\,mHz for different SN~Ia rates. Solid lines represent the results of the Monte Carlo simulation where we combined uncertainties in the number count of SN~Ia progenitors (assumed to be Poisson-distributed) with LISA's ability to measure the chirp mass. Dashed lines show the number count error only. 
      }
      \label{fig:rate}
\end{figure}

%%%%%%%%%%%%%%%%%%%%%%%%%%%%%%%%

To estimate the error on ${\cal R}_{\rm SNIa}$, we first sampled $N_{\rm SNIa}(>f_{\rm GW})$ from a Poisson distribution with the mean set by assuming SN~Ia merger rates of $(2, 4, 7)\times 10^{-3}$ yr$^{-1}$ \citep[e.g.][]{cap01,li11,mao18}, derived based on EM observations. Our assumption on the underlying `true' merger rate sets the number of detected binaries, which we then distributed in frequency according to Eq.~\ref{eqn:dNdf}. As a reference, setting the SN~Ia merger rate to $2 \times 10^{-3}$ yr$^{-1}$ ($7 \times 10^{-3}$ yr$^{-1}$ ) yields 1069, 57, 15 (3743, 200, 51) binaries with a chirp mass of 0.74\,M$_\odot$ at frequencies above 2, 6, and 10\,mHz, respectively.
To mimic the effects of measurement uncertainties, each chirp mass measurement is subsequently drawn from a Gaussian distribution, the width of which is interpolated from our results for the posterior widths for the \(1.05\,\textrm{M}_{\odot} + 0.7\,\textrm{M}_{\odot}\) binary in Fig.~\ref{fig:Mchirp}. We performed Monte Carlo simulations over an arbitrarily large number of realisations. 
The result is illustrated in Fig.~\ref{fig:rate} (solid lines); for comparison, we also plot the number count error only (dashed lines). 
This experiment demonstrates that LISA's error on the merger rate including binaries at $f_{\rm GW}<4$\,mHz will be dominated by the large number of binaries with relatively poor constraints on the chirp mass ($\sim$10\%; cf. Fig.~\ref{fig:Mchirp}). These binaries are about an order of magnitude more numerous than those at $f_{\rm GW}>4$\,mHz,  where the chirp mass constraint improves to $\sim1$\% level and better. 
On the other hand, in the sample including binaries with $f_{\rm GW}>4$\,mHz, the error is primarily driven by the Poissonian error given the lower number of sources ($< 50 - 10$ at $f_{\rm GW}>10$\,mHz for the considered SN~Ia rates). The measurement is optimal for the subsample of binaries with frequencies of less than $4$\,mHz, where, based on the LISA observations alone, we achieve SN~Ia merger rate error below 10\%. It might also be possible to obtain an improved measurement of the rate ${\cal R}_{\rm SNIa}$ (i.e.\ with a smaller uncertainty) by also using the larger, but incomplete, sample of binaries with lower frequencies, although this would require a more detailed population-level, hierarchical analysis that accounts for selection effects.

If the merger rate of WD+WD LISA sources inferred in this way turns out to be consistent with the known rate of SNe~Ia, this would be persuasive evidence in favour of the double-degenerate scenario. If, on the other hand, the rates do not match, then this suggests that at least some fraction of SNe~Ia come from another channel.

%--------------------------------------------------------------------
\section{The chance of a direct observation of a final inspiral and SN~Ia} \label{sec:3}

As discussed in Sect.~\ref{sec:2}, the double-degenerate SN~Ia progenitor scenario can already be confirmed or rejected based on the number/merger rate of close, massive WD+WD binaries that LISA will detect across our Galaxy.
Given the observed SN~Ia rate in the Milky Way-like galaxies, $\mathcal{R}_{\rm SNIa}= (3 - 7)\times 10^{-3}$ yr$^{-1}$ \citep[e.g.][]{cap01,li11,mao18}, we expect a 3 -- 7\% chance of a SN~Ia event during LISA's maximum lifetime of 10\,yr. Additionally, the most massive SN~Ia WD+WD progenitors LISA can also be observed in the Magellanic Clouds \citep{kor20, roe20,kei23} and possibly as far as the Andromeda galaxy \citep{kor18}. Here we outline the implications if such an opportunity arises. In Sect.~\ref{sec:GW_signal} we illustrate the expected GW signal, assuming a double-degenerate scenario in which the primary white dwarf undergoes double detonation \citep{liv90,fin10}.
In Sect.~\ref{sec:multi-messenger} we briefly discuss the potential insights that could be gained from multi-messenger observations.

\subsection{Gravitational wave signal} \label{sec:GW_signal}

  \begin{figure*}
   \centering
   \includegraphics[width=0.9\textwidth]{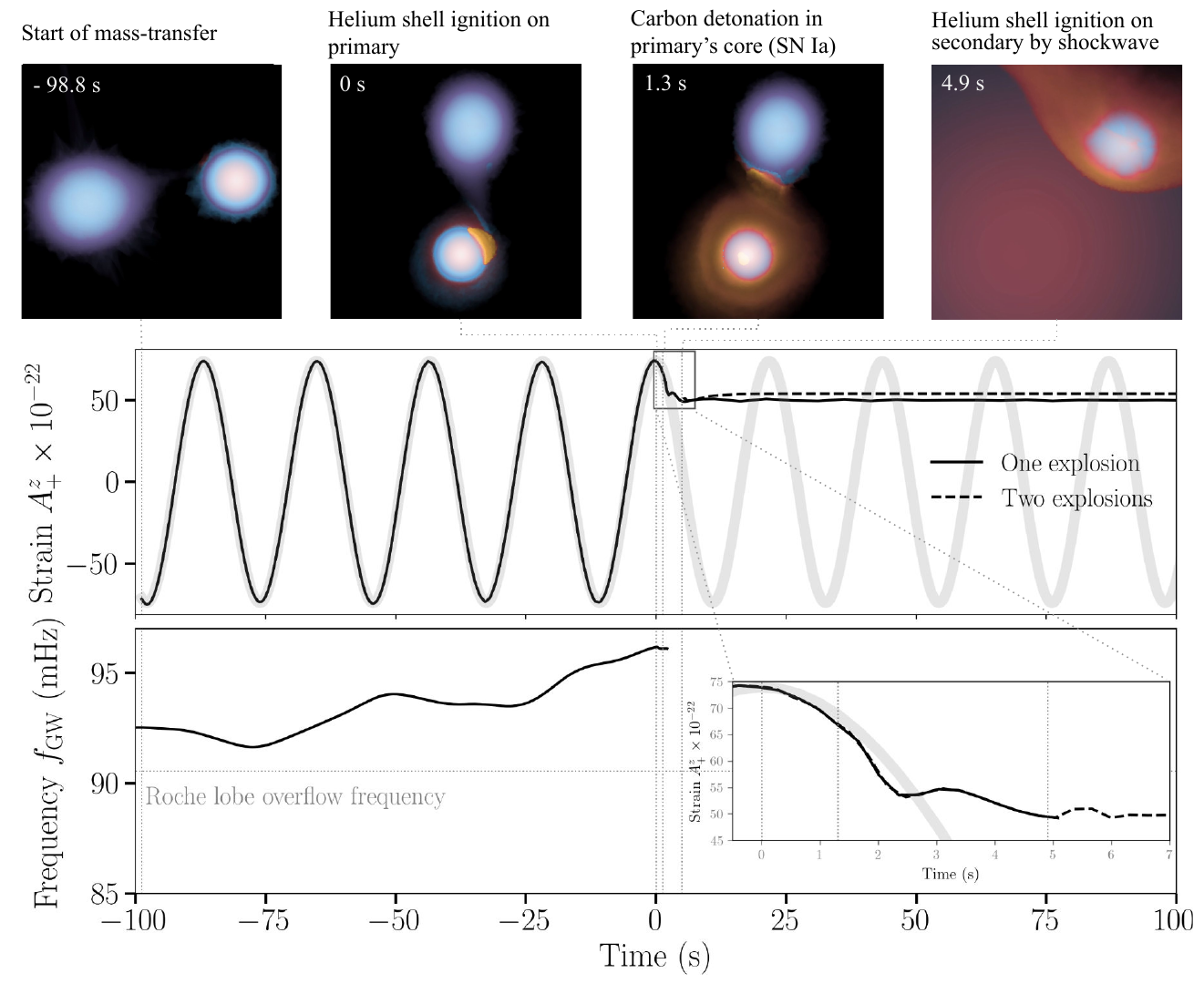}
      \caption{GW frequency and strain as a function of time derived from the 3D hydrodynamical simulation of the 1.05\,M$_\odot$ and 0.7\,M$_\odot$ carbon-oxygen white dwarf binary in \citet{pak22}, in which the primary undergoes double detonation. 
      {\it Top panels:} Key steps of the double-detonation mechanism. $t=-98.8$\,s is the onset of the mass transfer from the secondary (the visibly larger, fluffier white dwarf) onto the primary (the more compact of the two). $t=0$\,s is the helium-shell ignition on the primary. At  $t=1.3$\,s the helium detonation wraps around the primary, triggering the detonation of the primary's carbon-oxygen core.\ At $t=4.9$\,s, the carbon detonation completely burns  the primary (resulting in a SN~Ia) and triggers helium-shell ignition on the secondary. 
      {\it Middle panel:} Evolution of GW strain.\ The $A_+^z$ GW polarisation amplitude is computed based on the output of the hydrodynamical simulation following the \citet{mor23a} method and assuming a distance of 8.2\,kpc. The solid black line shows the `one-explosion' model, in which only the primary undergoes double detonation; the dashed line represents the `two-explosion' model, in which the explosion of the primary also triggers a double detonation of the secondary. In both cases, after the SN~Ia explosion, the amplitude levels off at a constant value until the end of the simulation, which represents the homogeneous expansion of the supernova ejecta. For comparison, the thick grey-shaded line represents a monochromatic signal of 92 Hz frequency and of equivalent amplitude.  
      The inset shows a zoomed-in view between -0.5\,s and 7\,s. 
      {\it Bottom panel:} Evolution of the GW frequency (solid line) that we computed directly from the binary separation obtained from the hydrodynamical simulation.}
      \label{fig:expl}
   \end{figure*}

To illustrate the GW signal from a WD+WD binary resulting in a SN~Ia, we adopted a 3D hydrodynamical simulation of \citet{pak22}.  
The simulation represents the last few orbits of a 1.05\,M$_\odot$ and 0.7\,M$_\odot$ carbon-oxygen WD+WD binary system, in which the primary undergoes the double-detonation mechanism via a dynamical ignition on the surface of the primary \citep{gui10,pak13,pak22,raj24}. 

We computed the GW signal following the same methods as in \citet[][see also \citealt{sei15}]{mor23a}. The amplitude of the GW signal is derived from the approximate quadrupole radiation from Newtonian gravity, following the numerical approach described in \citet{bla90}. We summarise the main points of this computation in Appendix~\ref{sec:appendix-gw}. In Fig.~\ref{fig:expl} we show the obtained evolution of GW frequency and strain amplitude as a function of time, assuming a distance of 8.2\,kpc as in Sect.~\ref{sec:Mchirp}. In the figure, we consider the most optimistic scenario for detectability, corresponding to a face-on binary orientation. This means the plane of the binary's orbit is perpendicular to the line of sight, maximising the `+' GW polarisation component (see Appendix~\ref{sec:appendix-gw}), which is shown in the middle panel of the figure. We estimate the time evolution of the GW frequency using $f_{\mathrm{GW}} = 2/P_{\mathrm{orb}}$, with $P_{\mathrm{orb}}$ taken directly from the output of the hydrodynamical simulation.

The first part of the signal (between -100\,s and 0\,s) represents the end of the inspiral phase starting with the onset of mass-transfer from the secondary (0.7\,M$_\odot$) onto the primary (1.05\,M$_\odot$) white dwarf via Roche-lobe overflow. During these 100\,s, the primary gains about $10^{-2}$\,M$_\odot$ of $^4$He. The evolution of the binary in this phase can still be described as in Sect.~\ref{sec:2}; indeed, the GW signal shows no significant deviation from a regular sinusoidal waveform. During this time, the orbital separation changes only slightly before the binary is disrupted, resulting in a small increase in GW frequency from 92\,Hz to\,96 Hz (bottom panel).

At $t=0$, the helium detonation ignites close to the point where the accretion stream hits the surface of the primary. The helium detonation then wraps around the primary burning its helium shell and sending a shockwave into its core. This shockwave converges in a single point in the core of the primary white dwarf and ignites a carbon detonation there at $t=1.3$\,s. The carbon detonation completely burns and destroys the primary white dwarf ($t=4.9$\,s), generating a SN~Ia. After the detonation of the primary, the strain amplitude in Fig.~\ref{fig:expl} shows a small excess signal (see the inset), which then levels off at a constant value until the end of the simulation $t=100$\,s. This part of GW signal is generated by the homogeneous expansion of the supernova ejecta.

At the same time, the shockwave of the explosion of the primary hits the secondary white dwarf. It ignites the helium shell of the secondary at around $t=2$\,s and additionally sends a shockwave into its core directly. This shockwave, supported by the helium detonation burning the remaining helium around the secondary white dwarf converges in its core at $t=4.9$\,s. The converging shockwave in the secondary white dwarf fails to ignite a carbon detonation in its core and the secondary white dwarf survives until the end of the simulation 100\,s later (the `one-explosion' model in Fig.~\ref{fig:expl}). 
It is interesting to note that a detonation initiation in the secondary is physically plausible, given its density and temperature at $t=4.9$\,s. We show the GW signal for this possible scenario generated by artificially igniting a carbon detonation in the secondary at the convergence point of the shockwave  (the `two-explosion' model in Fig.~\ref{fig:expl}). In this case, the second carbon detonation burns the secondary completely within about 1\,s and destroys it as well. 
In the `two-explosion' model, we observe an additional small excess signal between 5s and 6s in Fig.~\ref{fig:expl}, with the strain amplitude levelling off at a slightly higher constant value than in the `one-explosion' model due to the expansion of the secondary's ejecta. At the time of writing, the likelihood of the secondary’s detonation remains uncertain, and it is unclear whether the two models (one- and two-explosion) can be differentiated through EM observations \citep[e.g.][]{pak22,pol24}. Given the small additional strain amplitude produced as a result of the secondary's explosion, it is unlikely that this extra signal can be confidently extracted from the actual GW data. A more detailed technical investigation would be required to fully explore this possibility.

In summary, Fig.~\ref{fig:expl} illustrates that the binary emits nearly monochromatic GWs until just before the primary is disrupted by carbon detonation in the core, at which point the system virtually ceases to emit GWs. 
If the double-detonation mechanism is responsible for a significant portion of normal SNe~Ia (resulting from a detonation of $0.8 - 1.1$\,M$_\odot$ primary), the ignition and explosion must occur prior to the merger, leading to a `clean' disappearance of the binary's GW signal as in our example. 
If, on the other hand, the explosion occurs after the merger, such as in the case of the lower luminosity thermonuclear transient than a normal SN~Ia (e.g. resulting from the merger of a 0.6\,M$_\odot$ carbon-oxygen white dwarf with a 0.4\,M$_\odot$ helium white dwarf; see \citealt{mor24}) the GW signal will comprise both the merger and post-merger phases until the double detonation occurs \citep[see also][]{dan11,seto23}. In this case, the delay between the merger and the explosion is of the order of several minutes. A more detailed description of this scenario will be provided in a dedicated study.

\subsection{Chance of a multi-messenger observation} \label{sec:multi-messenger}

If a SN~Ia occurs in the Milky Way while LISA is operational, we might observe a GW signal followed by the detection of EM signal \citep[e.g. see synthetic light curves and spectra in][]{pak22}, followed by a shut off of the GW signal. 
Such an event would directly confirm the double-degenerate scenario, particularly demonstrating that SNe~Ia are rapid dynamical explosions, unfolding over just a few seconds, as suggested by recent double-detonation models (cf. Sect.~\ref{sec:GW_signal}) or violent mergers \citep[e.g.][]{pak10,pak12}. Any measurable delay between the GW and EM signals could indicate alternative detonation mechanisms and would represent an equally significant discovery.

A constraint on the progenitor binary component masses from the inspiral phase of the GW signal from LISA (see Sect.~\ref{sec:Mchirp}) and the observation of SNe~Ia would lead to significant progress in modelling explosion physics of SNe~Ia, even from a single observed event. Combining it with the EM signal would link the initial pre-explosion state of the system, and the outcome of the explosion in the form of SN~Ia ejecta and an EM display. This would likely allow us to constrain and improve on the inherent uncertainties in the modelling of explosion physics for SNe~Ia. 

Observing a GW signal without a corresponding EM signal would also provide valuable insights, leading to several possible interpretations. First, with LISA's capability to localise WD+WD binaries within an area $\ll$1\,deg$^2$ \citep[e.g.][]{fin23}, it could allow us to confirm or exclude that no EM emission is detectable due to strong extinction. 
If extinction is ruled out, the absence of EM emission might suggest a traditional double-degenerate super-Chandrasekhar scenario (i.e. a WD+WD merger with a total mass exceeding Chandrasekhar mass limit) with a long delay that may extend up to $10^4$\,yr \citep[e.g.][]{nomoto85,shen2012}. 

Finally, if an EM event is observed without a corresponding GW signal in the LISA band, this could indicate a merger with a delay-time between merger and explosion longer than the observing time of LISA (as discussed above) or a single-degenerate scenario, in which case the GW signal in the LISA band is not expected. 
Studies by \citet{fal11} and \citet{sei15} computed GW signals for SNe~Ia produced via different single-degenerate models. While the progenitor systems (carbon-oxygen white dwarf + main sequence or a red giant) would emit at frequencies below $10^{-4}\,$Hz, these studies find that the asymmetric explosion of the white dwarf would generate a short-duration GW signal between $\sim$0.4\,Hz and $\sim$2.5\,Hz, the exact frequency depends on the specifics of the model.

In the fortunate event that we observe a SN~Ia, this multi-messenger approach, leveraging both GW and EM information, promises  significant progress in our understanding of SNe~Ia, potentially directly resolving longstanding questions about their progenitors and providing crucial insights into the explosion mechanism. 

\subsection{Expectations for neutrino emission} \label{sec:neutrinos}

Neutrinos represent yet another messenger that could offer valuable insights into supernovae \citep[e.g.][]{jan17}. For example, neutrinos from a core-collapse supernova were observed in 1987 \citep[e.g.][]{hir1987}. Despite the very limited signal, they provided valuable insights into multiple neutrino properties and helped test the fundamental principles of the core-collapse explosion mechanism. Should we detect a neutrino signal from a SN~Ia, it would offer direct insights into the explosion mechanism \citep[][]{nom93,odr11,wri16,wri17}.

In SNe~Ia, neutrinos are mainly produced in the first second immediately after thermonuclear burning. They are produced primarily via two processes: electron captures on protons, neutrons, and nuclei (mainly $^{12}$C and $^{16}$O for white dwarfs that can undergo SNe~Ia), and thermal production (e.g. electron-positron annihilation, bremsstrahlung, and recombination processes). Compared to core-collapse (type II) supernovae, however, both neutrino production channels are weaker by many orders of magnitude. In the pre-explosion phase, the typical densities of white dwarfs are not high enough to overcome the threshold energy for electron capture reactions. These conditions can only be met for a short duration (of the order of couple of seconds) during thermonuclear burning. Thermal neutrino production is also lower because the temperatures in SN~Ia explosions are lower compared to those in the core-collapse supernovae ($10^9$\,K vs $10^{11}$\,K). In fact, considering a near-Chandrasekhar mass SN~Ia model \citet{sei15} obtained a total neutrino luminosity (generated by both types of neutrino production processes) of the order of $10^{49}$\,erg\,s$^{-1}$, while core-collapse supernovae produce an energy flux of the order of $10^{53}$\,erg\,s$^{-1}$ over 10\,s.

Given the weaker neutrino production in SNe~Ia, the detection prospects are significantly lower compared to core-collapse supernovae. \citet{wri16} calculated interaction rates for various detectors, including the Super-Kamiokande (Super-K; \citealt{Super-K}), future Hyper-Kamiokande (Hyper-K; \citealt{Hyper-K}), Jiangmen Underground Neutrino Observatory (JUNO; \citealt{JUNO}), Deep Underground Neutrino Experiment (DUNE; \citealt{DUNE}), and IceCube Neutrino Observatory (IceCube; \citealt{IceCube}). They found that at a distance of 10\,kpc, which is a likely distance due to the peak of stellar density near the Galactic centre at $\sim$8\,kpc, the expected number of neutrino interactions is very low. Specifically, Super-K, JUNO, and DUNE are expected to detect only a few events, while Hyper-K may observe several tens of events. At 1\,kpc, a distance at which a SN~Ia is much less likely, JUNO, Super-K, and DUNE are expected to register a few events, while IceCube and Hyper-K would register several tens of events.

The neutrino calculations discussed above have been performed for single-degenerate near-Chandrasekhar mass SN~Ia models. In this scenario, it has even been suggested that based on neutrino emission one would be able to differentiate between different detonation mechanisms \citep{wri17}. However, given lower densities and temperatures in the sub-Chandrasekhar double-detonation scenario considered in this study, we expect neutrino emission and prospects for detection to be even lower, of the order of $10^{45}$\,erg\,s$^{-1}$. In fact, detecting neutrinos from a SN~Ia would be a strong indication of explosive nuclear burning at densities above $10^9$\,g\,cm$^{-3}$, which would suggest a near-Chandrasekhar mass white dwarf undergoing deflagration \citep{sei15}. A non-detection of a neutrino signal, on the other hand, would favour models involving detonations in less massive white dwarfs, such as the violent merger or double-detonation models.

\section{Summary and conclusions} \label{sec:4}

The nature of SNe~Ia has long been one of the most debated open questions in astrophysics. In this study, we reviewed the expected insights that GW observations with LISA will provide about these transient phenomena. In particular, we argue that GW observations provide the most promising way for testing the so-called double-degenerate progenitor scenario, which involves a binary comprising two white dwarfs that either interact with each other or merge via GW radiation, resulting in a SN~Ia. LISA's ability to survey WD+WD binaries across the entirety of the Milky Way makes it an ideal tool for testing this progenitor scenario. 

In addition to LISA's science objectives outlined in \citet{Redbook}, we provide some quantitative estimates on the science that LISA will likely deliver in the context of SNe~Ia:
\begin{itemize}
    \item A complete sample of Galactic WD+WD binaries with GW frequencies above $2-3\,$mHz, corresponding to orbital periods shorter than 16 to 11 minutes. This implies that by simply counting WD+WD binaries in the LISA sample -- without any concerns regarding selection effects -- we will be able to estimate whether there are enough double-degenerate binaries in the Milky Way to account for the observed SN~Ia merger rate.
    \item Constraints on the binary chirp mass to better than 1\% at frequencies of 4--6 mHz. This, in turn, will allow us to set a lower bound on the primary mass and to differentiate which WD+WD binaries in the LISA's sample will eventually result in a SN~Ia ($m_1>0.8$\,M$_\odot$).
    \item Constraints on the merger rate of WD+WD binaries in the Milky Way to better than 4--9\% (depending on the true WD+WD merger rate). If this turns out to be consistent with the SN~Ia rate measured based on EM observations in Milky Way-like galaxies, it would be persuasive evidence in favour of the double-degenerate scenario.
\end{itemize}

We also discussed the possibility of a direct observation of a final inspiral and SN~Ia event; based on observational estimates, the likelihood of this is 3--7\% for an extended LISA mission duration of 10\,yr. We have illustrated the expected GW signal for a double-degenerate scenario in which the primary white dwarf undergoes double detonation. In this case, the ignition and explosion of the primary occur prior to the merger, leading to a ‘clean’ disappearance of the binary’s quasi-monochromatic GW signal. We speculate that even a single multi-messenger detection would significantly advance the modelling of SN~Ia explosions, reducing the parameter space on the initial conditions (based on constraints from the GW inspiral) and final EM ejecta. If we observe a quasi-monochromatic GW signal around 90\,mHz -- the exact frequency depends on the binary -- followed by the detection of an EM signal and a subsequent shut off of the GW signal, we would confirm the double-degenerate scenario in which a SN~Ia results from a double detonation of the primary white dwarf.

\begin{acknowledgements}
We thank Monica Colpi and Gijs Nelemans, who inspired this study while working on the LISA Definition Study Report (Red Book). Additionally, we thank Carlos Badenes, Stéphane Blondin, Jakob Hein, Stephen Justham, Antoine Klein, Hannah Middleton, Abinaya Swaruba Rajamuthukumar, Alberto Sesana, Nicola Tamanini, Silvia Toonen, and Alberto Vecchio for their useful discussions and suggestions for this study. We also thank the anonymous referee a constrictive report, which helped us to improve the manuscript.\\
%\vk{This research made use of \textsc{matplotlib} \citep{Hunter_2007}, \textsc{numpy} \citep{Numpy_2006, Numpy_2011}, and \textsc{scipy} \citep{Virtanen_2020} \textsc{python} packages.}

\end{acknowledgements}

\bibliographystyle{aa} % style aa.bst
\bibliography{typeIa_biblio} % your references Yourfile.bib

\begin{thebibliography}{91}
\expandafter\ifx\csname natexlab\endcsname\relax\def\natexlab#1{#1}\fi

\bibitem[{Aartsen {et~al.}(2017)}]{IceCube}
Aartsen, M.~G. {et~al.} 2017, Journal of Instrumentation, 12, P03012

\bibitem[{{Abi} {et~al.}(2020){Abi}, {Acciarri}, {Acero}, {Adamov}, {Adams}, {Adinolfi}, {Ahmad}, {Ahmed}, {Alion}, {Alonso Monsalve}, \& et~al.}]{DUNE}
{Abi}, B., {Acciarri}, R., {Acero}, M.~A., {et~al.} 2020, arXiv e-prints, arXiv:2002.03005

\bibitem[{{Ajith} {et~al.}(2024){Ajith}, {Amaro Seoane}, {Arca Sedda}, {Arcodia}, {Badaracco}, {Belgacem}, {Benetti}, {Bobrick}, {Bonforte}, {Bortolas}, {Braito}, {Branchesi}, {Burrows}, {Cappellaro}, {Della Ceca}, {Chakraborty}, {Chalathadka Subrahmanya}, {Coughlin}, {Covino}, {Derdzinski}, {Doshi}, {Falanga}, {Foffa}, {Franchini}, {Frigeri}, {Futaana}, {Gerberding}, {Gill}, {Di Giovanni}, {Giudice}, {Giustini}, {Gl{\"a}ser}, {Harms}, {van Heijningen}, {Iacovelli}, {Kavanagh}, {Kawamura}, {Kenath}, {Keppler}, {Kobayashi}, {Komatsu}, {Korol}, {Krishnendu}, {Kumar}, {Longo}, {Maggiore}, {Mancarella}, {Maselli}, {Mastrobuono-Battisti}, {Mazzarini}, {Melandri}, {Melini}, {Menina}, {Miniutti}, {Mitra}, {Mor{\'a}n-Fraile}, {Mukherjee}, {Muttoni}, {Olivieri}, {Onori}, {Alessandra Papa}, {Patat}, {Piran}, {Piranomonte}, {Roper Pol}, {Pookkillath}, {Prasad}, {Prasad}, {De Rosa}, {Chowdhury}, {Serafinelli}, {Sesana}, {Severgnini}, {Stallone}, {Tissino}, {Tkal{\v{c}}i{\'c}}, {Tomasella}, {Toscani}, {Vartanyan},
  {Vignali}, {Zaccarelli}, {Zeoli}, \& {Zuccarello}}]{LGWAastro}
{Ajith}, P., {Amaro Seoane}, P., {Arca Sedda}, M., {et~al.} 2024, arXiv e-prints, arXiv:2404.09181

\bibitem[{An {et~al.}(2016)}]{JUNO}
An, F. {et~al.} 2016, Journal of Physics G: Nuclear and Particle Physics, 43, 030401

\bibitem[{{Badenes} \& {Maoz}(2012)}]{badenes12}
{Badenes}, C. \& {Maoz}, D. 2012, \apjl, 749, L11

\bibitem[{{Blanchet} {et~al.}(1990){Blanchet}, {Damour}, \& {Schaefer}}]{bla90}
{Blanchet}, L., {Damour}, T., \& {Schaefer}, G. 1990, \mnras, 242, 289

\bibitem[{{Branchesi} {et~al.}(2023){Branchesi}, {Falanga}, {Harms}, {Jani}, {Katsanevas}, {Lognonn{\'e}}, {Badaracco}, {Cacciapuoti}, {Cappellaro}, {Dell'Agnello}, {de Raucourt}, {Frigeri}, {Giardini}, {Jennrich}, {Kawamura}, {Korol}, {Landr{\o}}, {Majstorovi{\'c}}, {Marmat}, {Mazzali}, {Muccino}, {Patat}, {Pian}, {Piran}, {Rosat}, {Rowan}, {St{\"a}hler}, \& {Tissino}}]{bra23}
{Branchesi}, M., {Falanga}, M., {Harms}, J., {et~al.} 2023, \ssr, 219, 67

\bibitem[{{Breivik} {et~al.}(2020){Breivik}, {Coughlin}, {Zevin}, {Rodriguez}, {Kremer}, {Ye}, {Andrews}, {Kurkowski}, {Digman}, {Larson}, \& {Rasio}}]{bre20}
{Breivik}, K., {Coughlin}, S., {Zevin}, M., {et~al.} 2020, \apj, 898, 71

\bibitem[{{Buscicchio} {et~al.}(2021){Buscicchio}, {Klein}, {Roebber}, {Moore}, {Gerosa}, {Finch}, \& {Vecchio}}]{bus21}
{Buscicchio}, R., {Klein}, A., {Roebber}, E., {et~al.} 2021, \prd, 104, 044065

\bibitem[{{Buscicchio} {et~al.}(2019){Buscicchio}, {Roebber}, {Goldstein}, \& {Moore}}]{2019PhRvD.100h4041B}
{Buscicchio}, R., {Roebber}, E., {Goldstein}, J.~M., \& {Moore}, C.~J. 2019, \prd, 100, 084041

\bibitem[{{Cappellaro} \& {Turatto}(2001)}]{cap01}
{Cappellaro}, E. \& {Turatto}, M. 2001, in Astrophysics and Space Science Library, Vol. 264, The Influence of Binaries on Stellar Population Studies, ed. D.~{Vanbeveren}, 199

\bibitem[{{Colpi} {et~al.}(2024){Colpi}, {Danzmann}, {Hewitson}, {Holley-Bockelmann}, {Jetzer}, {Nelemans}, {Petiteau}, {Shoemaker}, {Sopuerta}, {Stebbins}, {Tanvir}, {Ward}, {Weber}, {Thorpe}, {Daurskikh}, {Deep}, {Fern{\'a}ndez N{\'u}{\~n}ez}, {Garc{\'\i}a Marirrodriga}, {Gehler}, {Halain}, {Jennrich}, {Lammers}, {Larra{\~n}aga}, {Lieser}, {L{\"u}tzgendorf}, {Martens}, {Mondin}, {Piris Ni{\~n}o}, {Amaro-Seoane}, {Arca Sedda}, {Auclair}, {Babak}, {Baghi}, {Baibhav}, {Baker}, {Bayle}, {Berry}, {Berti}, {Boileau}, {Bonetti}, {Brito}, {Buscicchio}, {Calcagni}, {Capelo}, {Caprini}, {Caputo}, {Castelli}, {Chen}, {Chen}, {Chua}, {Davies}, {Derdzinski}, {Domcke}, {Doneva}, {Dvorkin}, {Mar{\'\i}a Ezquiaga}, {Gair}, {Haiman}, {Harry}, {Hartwig}, {Hees}, {Heffernan}, {Husa}, {Izquierdo-Villalba}, {Karnesis}, {Klein}, {Korol}, {Korsakova}, {Kupfer}, {Laghi}, {Lamberts}, {Larson}, {Le Jeune}, {Lewicki}, {Littenberg}, {Madge}, {Mangiagli}, {Marsat}, {Vilchez}, {Maselli}, {Mathews}, {van de Meent}, {Muratore}, {Nardini},
  {Pani}, {Peloso}, {Pieroni}, {Pound}, {Quelquejay-Leclere}, {Ricciardone}, {Rossi}, {Sartirana}, {Savalle}, {Sberna}, {Sesana}, {Shoemaker}, {Slutsky}, {Sotiriou}, {Speri}, {Staab}, {Steer}, {Tamanini}, {Tasinato}, {Torrado}, {Torres-Orjuela}, {Toubiana}, {Vallisneri}, {Vecchio}, {Volonteri}, {Yagi}, \& {Zwick}}]{Redbook}
{Colpi}, M., {Danzmann}, K., {Hewitson}, M., {et~al.} 2024, arXiv e-prints, arXiv:2402.07571

\bibitem[{{Cutler}(1998)}]{cut98}
{Cutler}, C. 1998, \prd, 57, 7089

\bibitem[{{Dan} {et~al.}(2011){Dan}, {Rosswog}, {Guillochon}, \& {Ramirez-Ruiz}}]{dan11}
{Dan}, M., {Rosswog}, S., {Guillochon}, J., \& {Ramirez-Ruiz}, E. 2011, \apj, 737, 89

\bibitem[{{Falta} {et~al.}(2011){Falta}, {Fisher}, \& {Khanna}}]{fal11}
{Falta}, D., {Fisher}, R., \& {Khanna}, G. 2011, \prl, 106, 201103

\bibitem[{{Fiacco} {et~al.}(2024){Fiacco}, {Cornish}, \& {Yu}}]{fia24}
{Fiacco}, G., {Cornish}, N.~J., \& {Yu}, H. 2024, arXiv e-prints, arXiv:2405.10396

\bibitem[{{Filippenko}(1997)}]{fil97}
{Filippenko}, A.~V. 1997, \araa, 35, 309

\bibitem[{{Finch} {et~al.}(2023){Finch}, {Bartolucci}, {Chucherko}, {Patterson}, {Korol}, {Klein}, {Bandopadhyay}, {Middleton}, {Moore}, \& {Vecchio}}]{fin23}
{Finch}, E., {Bartolucci}, G., {Chucherko}, D., {et~al.} 2023, \mnras, 522, 5358

\bibitem[{{Fink} {et~al.}(2010){Fink}, {R{\"o}pke}, {Hillebrandt}, {Seitenzahl}, {Sim}, \& {Kromer}}]{fin10}
{Fink}, M., {R{\"o}pke}, F.~K., {Hillebrandt}, W., {et~al.} 2010, \aap, 514, A53

\bibitem[{Fukuda {et~al.}(2003)}]{Super-K}
Fukuda, Y. {et~al.} 2003, Nuclear Instruments and Methods in Physics Research A, 501, 418

\bibitem[{{GRAVITY Collaboration} {et~al.}(2021){GRAVITY Collaboration}, {Abuter}, {Amorim}, {Baub{\"o}ck}, {Berger}, {Bonnet}, {Brandner}, {Cl{\'e}net}, {Davies}, {de Zeeuw}, {Dexter}, {Dallilar}, {Drescher}, {Eckart}, {Eisenhauer}, {F{\"o}rster Schreiber}, {Garcia}, {Gao}, {Gendron}, {Genzel}, {Gillessen}, {Habibi}, {Haubois}, {Hei{\ss}el}, {Henning}, {Hippler}, {Horrobin}, {Jim{\'e}nez-Rosales}, {Jochum}, {Jocou}, {Kaufer}, {Kervella}, {Lacour}, {Lapeyr{\`e}re}, {Le Bouquin}, {L{\'e}na}, {Lutz}, {Nowak}, {Ott}, {Paumard}, {Perraut}, {Perrin}, {Pfuhl}, {Rabien}, {Rodr{\'\i}guez-Coira}, {Shangguan}, {Shimizu}, {Scheithauer}, {Stadler}, {Straub}, {Straubmeier}, {Sturm}, {Tacconi}, {Vincent}, {von Fellenberg}, {Waisberg}, {Widmann}, {Wieprecht}, {Wiezorrek}, {Woillez}, {Yazici}, {Young}, \& {Zins}}]{gravity21}
{GRAVITY Collaboration}, {Abuter}, R., {Amorim}, A., {et~al.} 2021, \aap, 647, A59

\bibitem[{{Guillochon} {et~al.}(2010){Guillochon}, {Dan}, {Ramirez-Ruiz}, \& {Rosswog}}]{gui10}
{Guillochon}, J., {Dan}, M., {Ramirez-Ruiz}, E., \& {Rosswog}, S. 2010, \apjl, 709, L64

\bibitem[{{Harms} {et~al.}(2021){Harms}, {Ambrosino}, {Angelini}, {Braito}, {Branchesi}, {Brocato}, {Cappellaro}, {Coccia}, {Coughlin}, {Della Ceca}, {Della Valle}, {Dionisio}, {Federico}, {Formisano}, {Frigeri}, {Grado}, {Izzo}, {Marcelli}, {Maselli}, {Olivieri}, {Pernechele}, {Possenti}, {Ronchini}, {Serafinelli}, {Severgnini}, {Agostini}, {Badaracco}, {Bertolini}, {Betti}, {Civitani}, {Collette}, {Covino}, {Dall'Osso}, {D'Avanzo}, {DeSalvo}, {Di Giovanni}, {Focardi}, {Giunchi}, {van Heijningen}, {Khetan}, {Melini}, {Mitri}, {Mow-Lowry}, {Naponiello}, {Noce}, {Oganesyan}, {Pace}, {Paik}, {Pajewski}, {Palazzi}, {Pallavicini}, {Pareschi}, {Pozzobon}, {Sharma}, {Spada}, {Stanga}, {Tagliaferri}, \& {Votta}}]{LGWA}
{Harms}, J., {Ambrosino}, F., {Angelini}, L., {et~al.} 2021, \apj, 910, 1

\bibitem[{Hirata {et~al.}(1987)}]{hir1987}
Hirata, K. {et~al.} 1987, Physical Review Letters, 58, 1490

\bibitem[{{Hoyle} \& {Fowler}(1960)}]{hoy60}
{Hoyle}, F. \& {Fowler}, W.~A. 1960, \apj, 132, 565

\bibitem[{{Huang} {et~al.}(2020){Huang}, {Hu}, {Korol}, {Li}, {Liang}, {Lu}, {Wang}, {Yu}, \& {Mei}}]{hua20}
{Huang}, S.-J., {Hu}, Y.-M., {Korol}, V., {et~al.} 2020, \prd, 102, 063021

\bibitem[{{Hyper-Kamiokande Proto-Collaboration} {et~al.}(2018){Hyper-Kamiokande Proto-Collaboration}, {:}, {Abe}, {Abe}, {Aihara}, {Aimi}, {Akutsu}, {Andreopoulos}, {Anghel}, {Anthony}, \& et~al.}]{Hyper-K}
{Hyper-Kamiokande Proto-Collaboration}, {:}, {Abe}, K., {et~al.} 2018, arXiv e-prints, arXiv:1805.04163

\bibitem[{{Iben} \& {Tutukov}(1984)}]{ibe84}
{Iben}, I., J. \& {Tutukov}, A.~V. 1984, \apjs, 54, 335

\bibitem[{{Janka}(2017)}]{jan17}
{Janka}, H.-T. 2017, in Handbook of Supernovae, ed. A.~W. {Alsabti} \& P.~{Murdin}, 1575

\bibitem[{{Karnesis} {et~al.}(2021){Karnesis}, {Babak}, {Pieroni}, {Cornish}, \& {Littenberg}}]{kar21}
{Karnesis}, N., {Babak}, S., {Pieroni}, M., {Cornish}, N., \& {Littenberg}, T. 2021, \prd, 104, 043019

\bibitem[{{Keim} {et~al.}(2023){Keim}, {Korol}, \& {Rossi}}]{kei23}
{Keim}, M.~A., {Korol}, V., \& {Rossi}, E.~M. 2023, \mnras, 521, 1088

\bibitem[{{Klein} {et~al.}(2022){Klein}, {Pratten}, {Buscicchio}, {Schmidt}, {Moore}, {Finch}, {Bonino}, {Thomas}, {Williams}, {Gerosa}, {McGee}, {Nicholl}, \& {Vecchio}}]{kle22}
{Klein}, A., {Pratten}, G., {Buscicchio}, R., {et~al.} 2022, arXiv e-prints, arXiv:2204.03423

\bibitem[{{Korol} {et~al.}(2022){Korol}, {Hallakoun}, {Toonen}, \& {Karnesis}}]{kor22}
{Korol}, V., {Hallakoun}, N., {Toonen}, S., \& {Karnesis}, N. 2022, \mnras, 511, 5936

\bibitem[{{Korol} {et~al.}(2018){Korol}, {Koop}, \& {Rossi}}]{kor18}
{Korol}, V., {Koop}, O., \& {Rossi}, E.~M. 2018, \apjl, 866, L20

\bibitem[{{Korol} {et~al.}(2017){Korol}, {Rossi}, {Groot}, {Nelemans}, {Toonen}, \& {Brown}}]{kor17}
{Korol}, V., {Rossi}, E.~M., {Groot}, P.~J., {et~al.} 2017, \mnras, 470, 1894

\bibitem[{{Korol} {et~al.}(2020){Korol}, {Toonen}, {Klein}, {Belokurov}, {Vincenzo}, {Buscicchio}, {Gerosa}, {Moore}, {Roebber}, {Rossi}, \& {Vecchio}}]{kor20}
{Korol}, V., {Toonen}, S., {Klein}, A., {et~al.} 2020, \aap, 638, A153

\bibitem[{{Lamberts} {et~al.}(2019){Lamberts}, {Blunt}, {Littenberg}, {Garrison-Kimmel}, {Kupfer}, \& {Sanderson}}]{lam19}
{Lamberts}, A., {Blunt}, S., {Littenberg}, T.~B., {et~al.} 2019, \mnras, 490, 5888

\bibitem[{{Li} {et~al.}(2011){Li}, {Chornock}, {Leaman}, {Filippenko}, {Poznanski}, {Wang}, {Ganeshalingam}, \& {Mannucci}}]{li11}
{Li}, W., {Chornock}, R., {Leaman}, J., {et~al.} 2011, \mnras, 412, 1473

\bibitem[{{Li} {et~al.}(2020){Li}, {Chen}, {Chen}, {Li}, {Yu}, \& {Han}}]{li20}
{Li}, Z., {Chen}, X., {Chen}, H.-L., {et~al.} 2020, \apj, 893, 2

\bibitem[{{Li} {et~al.}(2023){Li}, {Chen}, {Ge}, {Chen}, \& {Han}}]{li23}
{Li}, Z., {Chen}, X., {Ge}, H., {Chen}, H.-L., \& {Han}, Z. 2023, \aap, 669, A82

\bibitem[{{LISA Consortium Astrophysics Working Group} {et~al.}(2023){LISA Consortium Astrophysics Working Group}, {Amaro-Seoane}, {Andrews}, {Arca Sedda}, {Askar}, {Baghi}, {Balasov}, {Bartos}, {Bavera}, {Bellovary}, {Berry}, {Berti}, {Bianchi}, {Blecha}, {Blondin}, {Bogdanovi{\'c}}, {Boissier}, {Bonetti}, {Bonoli}, {Bortolas}, {Breivik}, {Capelo}, {Caramete}, {Cattorini}, {Charisi}, {Chaty}, {Chen}, {Chru{\'s}li{\'n}ska}, {Chua}, {Church}, {Colpi}, {D'Orazio}, {Danielski}, {Davies}, {Dayal}, {De Rosa}, {Derdzinski}, {Destounis}, {Dotti}, {Dutan}, {Dvorkin}, {Fabj}, {Foglizzo}, {Ford}, {Fouvry}, {Franchini}, {Fragos}, {Fryer}, {Gaspari}, {Gerosa}, {Graziani}, {Groot}, {Habouzit}, {Haggard}, {Haiman}, {Han}, {Istrate}, {Johansson}, {Khan}, {Kimpson}, {Kokkotas}, {Kong}, {Korol}, {Kremer}, {Kupfer}, {Lamberts}, {Larson}, {Lau}, {Liu}, {Lloyd-Ronning}, {Lodato}, {Lupi}, {Ma}, {Maccarone}, {Mandel}, {Mangiagli}, {Mapelli}, {Mathis}, {Mayer}, {McGee}, {McKernan}, {Miller}, {Mota}, {Mumpower}, {Nasim}, {Nelemans},
  {Noble}, {Pacucci}, {Panessa}, {Paschalidis}, {Pfister}, {Porquet}, {Quenby}, {Ricarte}, {R{\"o}pke}, {Regan}, {Rosswog}, {Ruiter}, {Ruiz}, {Runnoe}, {Schneider}, {Schnittman}, {Secunda}, {Sesana}, {Seto}, {Shao}, {Shapiro}, {Sopuerta}, {Stone}, {Suvorov}, {Tamanini}, {Tamfal}, {Tauris}, {Temmink}, {Tomsick}, {Toonen}, {Torres-Orjuela}, {Toscani}, {Tsokaros}, {Unal}, {V{\'a}zquez-Aceves}, {Valiante}, {van Putten}, {van Roestel}, {Vignali}, {Volonteri}, {Wu}, {Younsi}, {Yu}, {Zane}, {Zwick}, {Antonini}, {Baibhav}, {Barausse}, {Bonilla Rivera}, {Branchesi}, {Branduardi-Raymont}, {Burdge}, {Chakraborty}, {Cuadra}, {Dage}, {Davis}, {de Mink}, {Decarli}, {Doneva}, {Escoffier}, {Gandhi}, {Haardt}, {Lousto}, {Nissanke}, {Nordhaus}, {O'Shaughnessy}, {Portegies Zwart}, {Pound}, {Schussler}, {Sergijenko}, {Spallicci}, {Vernieri}, \& {Vigna-G{\'o}mez}}]{astrowp}
{LISA Consortium Astrophysics Working Group}, {Amaro-Seoane}, P., {Andrews}, J., {et~al.} 2023, Living Reviews in Relativity, 26, 2

\bibitem[{{LISA Consortium Waveform Working Group} {et~al.}(2023){LISA Consortium Waveform Working Group}, {Afshordi}, {Ak{\c{c}}ay}, {Amaro Seoane}, {Antonelli}, {Aurrekoetxea}, {Barack}, {Barausse}, {Benkel}, {Bernard}, {Bernuzzi}, {Berti}, {Bonetti}, {Bonga}, {Bozzola}, {Brito}, {Buonanno}, {C{\'a}rdenas-Avenda{\~n}o}, {Casals}, {Chernoff}, {Chua}, {Clough}, {Colleoni}, {Dhesi}, {Druart}, {Durkan}, {Faye}, {Ferguson}, {Field}, {Gabella}, {Garc{\'\i}a-Bellido}, {Gracia-Linares}, {Gerosa}, {Green}, {Haney}, {Hannam}, {Heffernan}, {Hinderer}, {Helfer}, {Hughes}, {Husa}, {Isoyama}, {Katz}, {Kavanagh}, {Khanna}, {Kidder}, {Korol}, {K{\"u}chler}, {Laguna}, {Larrouturou}, {Le Tiec}, {Leather}, {Lim}, {Lim}, {Littenberg}, {Long}, {Lousto}, {Lovelace}, {Lukes-Gerakopoulos}, {Lynch}, {Macedo}, {Markakis}, {Maggio}, {Mandel}, {Maselli}, {Mathews}, {Mourier}, {Neilsen}, {Nagar}, {Nichols}, {Nov{\'a}k}, {Okounkova}, {O'Shaughnessy}, {Oshita}, {O'Toole}, {Pan}, {Pani}, {Pappas}, {Paschalidis}, {Pfeiffer}, {Pompili},
  {Pound}, {Pratten}, {R{\"u}ter}, {Ruiz}, {Sam}, {Sberna}, {Shapiro}, {Shoemaker}, {Sopuerta}, {Spiers}, {Sundar}, {Tamanini}, {Thompson}, {Toubiana}, {Tsokaros}, {Upton}, {van de Meent}, {Vernieri}, {Wachter}, {Warburton}, {Wardell}, {Witek}, {Witzany}, {Yang}, {Zilh{\~a}o}, {Albertini}, {Arun}, {Bezares}, {Bonilla}, {Chapman-Bird}, {Cownden}, {Cunningham}, {Devitt}, {Dolan}, {Duque}, {Dyson}, {Fryer}, {Gair}, {Giacomazzo}, {Gupta}, {Han}, {Haas}, {Hirschmann}, {Huerta}, {Jetzer}, {Kelly}, {Khalil}, {Lewis}, {Lloyd-Ronning}, {Marsat}, {Nardini}, {Neef}, {Ottewill}, {Pantelidou}, {Piovano}, {Redondo-Yuste}, {Sagunski}, {Stein}, {Skoup{\'y}}, {Sperhake}, {Speri}, {Spieksma}, {Stevens}, {Trestini}, \& {Va{\~n}{\'o}-Vi{\~n}uales}}]{WaveformsWP}
{LISA Consortium Waveform Working Group}, {Afshordi}, N., {Ak{\c{c}}ay}, S., {et~al.} 2023, arXiv e-prints, arXiv:2311.01300

\bibitem[{{Liu} {et~al.}(2023){Liu}, {R{\"o}pke}, \& {Han}}]{liu23}
{Liu}, Z.-W., {R{\"o}pke}, F.~K., \& {Han}, Z. 2023, Research in Astronomy and Astrophysics, 23, 082001

\bibitem[{{Livio} \& {Mazzali}(2018)}]{liv18}
{Livio}, M. \& {Mazzali}, P. 2018, \physrep, 736, 1

\bibitem[{{Livne}(1990)}]{liv90}
{Livne}, E. 1990, \apjl, 354, L53

\bibitem[{{Luo} {et~al.}(2016){Luo}, {Chen}, {Duan}, {Gong}, {Hu}, {Ji}, {Liu}, {Mei}, {Milyukov}, {Sazhin}, {Shao}, {Toth}, {Tu}, {Wang}, {Wang}, {Yeh}, {Zhan}, {Zhang}, {Zharov}, \& {Zhou}}]{TianQin}
{Luo}, J., {Chen}, L.-S., {Duan}, H.-Z., {et~al.} 2016, Classical and Quantum Gravity, 33, 035010

\bibitem[{{Maoz} {et~al.}(2012){Maoz}, {Badenes}, \& {Bickerton}}]{maoz12}
{Maoz}, D., {Badenes}, C., \& {Bickerton}, S.~J. 2012, \apj, 751, 143

\bibitem[{{Maoz} {et~al.}(2018){Maoz}, {Hallakoun}, \& {Badenes}}]{mao18}
{Maoz}, D., {Hallakoun}, N., \& {Badenes}, C. 2018, \mnras, 476, 2584

\bibitem[{{Maoz} {et~al.}(2014){Maoz}, {Mannucci}, \& {Nelemans}}]{mao14}
{Maoz}, D., {Mannucci}, F., \& {Nelemans}, G. 2014, \araa, 52, 107

\bibitem[{{Moore} {et~al.}(2015){Moore}, {Cole}, \& {Berry}}]{moo15}
{Moore}, C.~J., {Cole}, R.~H., \& {Berry}, C.~P.~L. 2015, Classical and Quantum Gravity, 32, 015014

\bibitem[{{Mor{\'a}n-Fraile} {et~al.}(2024){Mor{\'a}n-Fraile}, {Holas}, {R{\"o}pke}, {Pakmor}, \& {Schneider}}]{mor24}
{Mor{\'a}n-Fraile}, J., {Holas}, A., {R{\"o}pke}, F.~K., {Pakmor}, R., \& {Schneider}, F. R.~N. 2024, \aap, 683, A44

\bibitem[{{Mor{\'a}n-Fraile} {et~al.}(2023){Mor{\'a}n-Fraile}, {Schneider}, {R{\"o}pke}, {Ohlmann}, {Pakmor}, {Soultanis}, \& {Bauswein}}]{mor23a}
{Mor{\'a}n-Fraile}, J., {Schneider}, F. R.~N., {R{\"o}pke}, F.~K., {et~al.} 2023, \aap, 672, A9

\bibitem[{{Munday} {et~al.}(2024){Munday}, {Pelisoli}, {Tremblay}, {Marsh}, {Nelemans}, {B{\'e}dard}, {Toonen}, {Breedt}, {Cunningham}, {O'Brien}, \& {Dawson}}]{mun24}
{Munday}, J., {Pelisoli}, I., {Tremblay}, P.~E., {et~al.} 2024, \mnras, 532, 2534

\bibitem[{{Nelemans} {et~al.}(2001){Nelemans}, {Yungelson}, {Portegies Zwart}, \& {Verbunt}}]{nel01}
{Nelemans}, G., {Yungelson}, L.~R., {Portegies Zwart}, S.~F., \& {Verbunt}, F. 2001, \aap, 365, 491

\bibitem[{{Nomoto} \& {Iben}(1985)}]{nomoto85}
{Nomoto}, K. \& {Iben}, I., J. 1985, \apj, 297, 531

\bibitem[{{Nomoto} {et~al.}(1993){Nomoto}, {Iwamoto}, {Tsujimoto}, \& {Hashimoto}}]{nom93}
{Nomoto}, K., {Iwamoto}, K., {Tsujimoto}, T., \& {Hashimoto}, M. 1993, in Frontiers of Neutrino Astrophysics, ed. Y.~{Suzuki} \& K.~{Nakamura}, 235--254

\bibitem[{{Nomoto} {et~al.}(1984){Nomoto}, {Thielemann}, \& {Yokoi}}]{non84}
{Nomoto}, K., {Thielemann}, F.~K., \& {Yokoi}, K. 1984, \apj, 286, 644

\bibitem[{{Odrzywolek} \& {Plewa}(2011)}]{odr11}
{Odrzywolek}, A. \& {Plewa}, T. 2011, \aap, 529, A156

\bibitem[{{Pakmor} {et~al.}(2022){Pakmor}, {Callan}, {Collins}, {de Mink}, {Holas}, {Kerzendorf}, {Kromer}, {Neunteufel}, {O'Brien}, {R{\"o}pke}, {Ruiter}, {Seitenzahl}, {Shingles}, {Sim}, \& {Taubenberger}}]{pak22}
{Pakmor}, R., {Callan}, F.~P., {Collins}, C.~E., {et~al.} 2022, \mnras, 517, 5260

\bibitem[{{Pakmor} {et~al.}(2010){Pakmor}, {Kromer}, {R{\"o}pke}, {Sim}, {Ruiter}, \& {Hillebrandt}}]{pak10}
{Pakmor}, R., {Kromer}, M., {R{\"o}pke}, F.~K., {et~al.} 2010, \nat, 463, 61

\bibitem[{{Pakmor} {et~al.}(2012){Pakmor}, {Kromer}, {Taubenberger}, {Sim}, {R{\"o}pke}, \& {Hillebrandt}}]{pak12}
{Pakmor}, R., {Kromer}, M., {Taubenberger}, S., {et~al.} 2012, \apjl, 747, L10

\bibitem[{{Pakmor} {et~al.}(2013){Pakmor}, {Kromer}, {Taubenberger}, \& {Springel}}]{pak13}
{Pakmor}, R., {Kromer}, M., {Taubenberger}, S., \& {Springel}, V. 2013, \apjl, 770, L8

\bibitem[{{Perlmutter} {et~al.}(1999){Perlmutter}, {Aldering}, {Goldhaber}, {Knop}, {Nugent}, {Castro}, {Deustua}, {Fabbro}, {Goobar}, {Groom}, {Hook}, {Kim}, {Kim}, {Lee}, {Nunes}, {Pain}, {Pennypacker}, {Quimby}, {Lidman}, {Ellis}, {Irwin}, {McMahon}, {Ruiz-Lapuente}, {Walton}, {Schaefer}, {Boyle}, {Filippenko}, {Matheson}, {Fruchter}, {Panagia}, {Newberg}, {Couch}, \& {Project}}]{per99}
{Perlmutter}, S., {Aldering}, G., {Goldhaber}, G., {et~al.} 1999, \apj, 517, 565

\bibitem[{{Peters}(1964)}]{pet64}
{Peters}, P.~C. 1964, Physical Review, 136, 1224

\bibitem[{{Pollin} {et~al.}(2024){Pollin}, {Sim}, {Pakmor}, {Callan}, {Collins}, {Shingles}, {R{\"o}pke}, \& {Srivastav}}]{pol24}
{Pollin}, J.~M., {Sim}, S.~A., {Pakmor}, R., {et~al.} 2024, \mnras, 533, 3036

\bibitem[{{Portegies Zwart} \& {Verbunt}(1996)}]{spz96}
{Portegies Zwart}, S.~F. \& {Verbunt}, F. 1996, \aap, 309, 179

\bibitem[{{Rajavel} {et~al.}(2024){Rajavel}, {Townsley}, \& {Shen}}]{raj24}
{Rajavel}, N., {Townsley}, D.~M., \& {Shen}, K.~J. 2024, arXiv e-prints, arXiv:2408.10981

\bibitem[{{Rebassa-Mansergas} {et~al.}(2019){Rebassa-Mansergas}, {Toonen}, {Korol}, \& {Torres}}]{reb19}
{Rebassa-Mansergas}, A., {Toonen}, S., {Korol}, V., \& {Torres}, S. 2019, \mnras, 482, 3656

\bibitem[{{Riess} {et~al.}(1998){Riess}, {Filippenko}, {Challis}, {Clocchiatti}, {Diercks}, {Garnavich}, {Gilliland}, {Hogan}, {Jha}, {Kirshner}, {Leibundgut}, {Phillips}, {Reiss}, {Schmidt}, {Schommer}, {Smith}, {Spyromilio}, {Stubbs}, {Suntzeff}, \& {Tonry}}]{rie98}
{Riess}, A.~G., {Filippenko}, A.~V., {Challis}, P., {et~al.} 1998, \aj, 116, 1009

\bibitem[{{Roebber} {et~al.}(2020){Roebber}, {Buscicchio}, {Vecchio}, {Moore}, {Klein}, {Korol}, {Toonen}, {Gerosa}, {Goldstein}, {Gaebel}, \& {Woods}}]{roe20}
{Roebber}, E., {Buscicchio}, R., {Vecchio}, A., {et~al.} 2020, \apjl, 894, L15

\bibitem[{{Ruan} {et~al.}(2018){Ruan}, {Guo}, {Cai}, \& {Zhang}}]{Taiji}
{Ruan}, W.-H., {Guo}, Z.-K., {Cai}, R.-G., \& {Zhang}, Y.-Z. 2018, arXiv e-prints, arXiv:1807.09495

\bibitem[{{Ruiter}(2020)}]{rui20}
{Ruiter}, A.~J. 2020, in White Dwarfs as Probes of Fundamental Physics: Tracers of Planetary, Stellar and Galactic Evolution, ed. M.~A. {Barstow}, S.~J. {Kleinman}, J.~L. {Provencal}, \& L.~{Ferrario}, Vol. 357, 1--15

\bibitem[{{Seitenzahl} {et~al.}(2015){Seitenzahl}, {Herzog}, {Ruiter}, {Marquardt}, {Ohlmann}, \& {R{\"o}pke}}]{sei15}
{Seitenzahl}, I.~R., {Herzog}, M., {Ruiter}, A.~J., {et~al.} 2015, \prd, 92, 124013

\bibitem[{{Sesana} {et~al.}(2020){Sesana}, {Lamberts}, \& {Petiteau}}]{ses20}
{Sesana}, A., {Lamberts}, A., \& {Petiteau}, A. 2020, \mnras, 494, L75

\bibitem[{{Seto}(2023)}]{seto23}
{Seto}, N. 2023, \mnras, 523, 577

\bibitem[{{Shah} \& {Nelemans}(2014)}]{sha14}
{Shah}, S. \& {Nelemans}, G. 2014, \apj, 791, 76

\bibitem[{{Shah} {et~al.}(2012){Shah}, {van der Sluys}, \& {Nelemans}}]{sha12}
{Shah}, S., {van der Sluys}, M., \& {Nelemans}, G. 2012, \aap, 544, A153

\bibitem[{{Shen}(2015)}]{shen15}
{Shen}, K.~J. 2015, \apjl, 805, L6

\bibitem[{{Shen} {et~al.}(2012){Shen}, {Bildsten}, {Kasen}, \& {Quataert}}]{shen2012}
{Shen}, K.~J., {Bildsten}, L., {Kasen}, D., \& {Quataert}, E. 2012, \apj, 748, 35

\bibitem[{{Shen} {et~al.}(2024){Shen}, {Boos}, \& {Townsley}}]{shen24}
{Shen}, K.~J., {Boos}, S.~J., \& {Townsley}, D.~M. 2024, arXiv e-prints, arXiv:2405.19417

\bibitem[{{Sim} {et~al.}(2010){Sim}, {R{\"o}pke}, {Hillebrandt}, {Kromer}, {Pakmor}, {Fink}, {Ruiter}, \& {Seitenzahl}}]{sim10}
{Sim}, S.~A., {R{\"o}pke}, F.~K., {Hillebrandt}, W., {et~al.} 2010, \apjl, 714, L52

\bibitem[{{Soker}(2024)}]{sok24}
{Soker}, N. 2024, The Open Journal of Astrophysics, 7, 31

\bibitem[{{Tang} {et~al.}(2024){Tang}, {Eldridge}, {Meyer}, {Lamberts}, {Boileau}, \& {van Zeist}}]{tang24}
{Tang}, P., {Eldridge}, J., {Meyer}, R., {et~al.} 2024, arXiv e-prints, arXiv:2405.20484

\bibitem[{{Thiele} {et~al.}(2023){Thiele}, {Breivik}, {Sanderson}, \& {Luger}}]{tie23}
{Thiele}, S., {Breivik}, K., {Sanderson}, R.~E., \& {Luger}, R. 2023, \apj, 945, 162

\bibitem[{{Toonen} {et~al.}(2012){Toonen}, {Nelemans}, \& {Portegies Zwart}}]{too12}
{Toonen}, S., {Nelemans}, G., \& {Portegies Zwart}, S. 2012, \aap, 546, A70

\bibitem[{{Toubiana} {et~al.}(2024){Toubiana}, {Karnesis}, {Lamberts}, \& {Miller}}]{tou24}
{Toubiana}, A., {Karnesis}, N., {Lamberts}, A., \& {Miller}, M.~C. 2024, arXiv e-prints, arXiv:2403.16867

\bibitem[{{Whelan} \& {Iben}(1973)}]{whe73}
{Whelan}, J. \& {Iben}, Icko, J. 1973, \apj, 186, 1007

\bibitem[{{Wilhelm} {et~al.}(2021){Wilhelm}, {Korol}, {Rossi}, \& {D'Onghia}}]{wil21}
{Wilhelm}, M. J.~C., {Korol}, V., {Rossi}, E.~M., \& {D'Onghia}, E. 2021, \mnras, 500, 4958

\bibitem[{{Wolz} {et~al.}(2021){Wolz}, {Yagi}, {Anderson}, \& {Taylor}}]{wol21}
{Wolz}, A., {Yagi}, K., {Anderson}, N., \& {Taylor}, A.~J. 2021, \mnras, 500, L52

\bibitem[{{Wright} {et~al.}(2017){Wright}, {Kneller}, {Ohlmann}, {R{\"o}pke}, {Scholberg}, \& {Seitenzahl}}]{wri17}
{Wright}, W.~P., {Kneller}, J.~P., {Ohlmann}, S.~T., {et~al.} 2017, \prd, 95, 043006

\bibitem[{{Wright} {et~al.}(2016){Wright}, {Nagaraj}, {Kneller}, {Scholberg}, \& {Seitenzahl}}]{wri16}
{Wright}, W.~P., {Nagaraj}, G., {Kneller}, J.~P., {Scholberg}, K., \& {Seitenzahl}, I.~R. 2016, \prd, 94, 025026

\end{thebibliography}

\begin{appendix}
\section{Computation of the gravitational wave signal}
\label{sec:appendix-gw}

Here we briefly summarise the approach for computing the GW signal based on the output of the hydrodynamic simulation as detailed in \citet{sei15} and \citet{mor23a}. The amplitude of the GWs can be derived from the second time derivative of the quadrupole moment, following a numerical approach based on the approximate quadrupole formula derived in \citet{bla90}. 

The gravitational quadrupole radiation field in the transverse-traceless gauge \( h_{ij}^{TT} \) can be written as
\begin{equation}
h_{ij}^{TT}(\textbf{x}, t) = \frac{2G}{c^4d} P_{ijkl}(\textbf{n}) \int \rho (2v_k v_l - x_k \partial_l \Phi - x_l \partial_k \Phi) \, {\rm d}^3x,
\end{equation}
where $P_{ijkl}(\textbf{n}) = (\delta_{ij} - n_{i}n_{k})(\delta_{jl}-n_{j}n_{k})-\frac{1}{2}(\delta_{ij}-n_in_j)(\delta_{kl}-n_k n_l)$ is the transverse-traceless projection operator, with $\textbf{n}=\textbf{x}/d$ being the normalised position vector, \( \rho \) is the density, \( \textbf{v} \) is the velocity, \( \partial_i \) denotes a partial derivative with respect to the spatial coordinate \( i \), and \( \Phi \) is the Newtonian gravitational potential; as before, \( G \) is the gravitational constant, \( c \) is the speed of light, and $d = |{\bf x}|$ is the distance to the source.

The amplitude of the GW radiation field can be written in terms of the two unit linear polarisation tensors \( e_+ \) and \( e_\times \), and for a chosen line of sight, the GW strain takes the form
\begin{equation}
h_{ij}^{TT}({\bf x}, t) = \frac{1}{d} (A_+ e^+_{ij} + A_\times e^\times_{ij}),
\end{equation}
where \( A_+ \) and \( A_\times \) are the amplitudes of the two GW polarisation at the source. 

As an example, we considered the most optimistic scenario for detectability, corresponding to a face-on binary orientation. This means that the orbital plane of the binary is perpendicular to the line of sight, which we assumed to be in the $z$-direction. In this case, the polarisation amplitudes are given by\begin{equation}
A_z^+ = A_{xx} - A_{yy}
\end{equation}
and
\begin{equation}
A_z^\times = 2A_{xy}
,\end{equation}
with \( A_{ij} \) defined as\begin{equation}
A_{ij} = \frac{G}{c^4} \int  \rho (2v_i v_j - x_i \partial_j \Phi - x_j \partial_i \Phi) \,{\rm d}^3x.
\end{equation}
We represent $A_z^+$ in Fig.~\ref{fig:expl} to provide a visual example that maximises the detectability of the GW signal. For different binary inclination angles, the relative amplitudes of the polarisation would vary, affecting the observed signal by a factor of a few; examples of the impact of the inclination on the signal-to-noise ratio for edge-on ($\cos \iota^{\rm inj} =0$) and face-on ($\cos \iota^{\rm inj} =1$) cases are presented in Table~\ref{tab:1} (see also \citealt{sha12, fin23}). Finally, we note that the GW amplitudes discussed here represent an ideal case, as they need to be combined with the detector's response functions, as detailed in \citet{cut98}.

\end{appendix}
\end{document}